\documentclass[11pt]{article}
\usepackage[right=1in,left=1in,top=2.0in,bottom=1in]{geometry}
\usepackage{latexsym,amsmath,amssymb,amsthm,amsfonts,bbm}
\usepackage{eurosym,ulem,sectsty,comment,array}
\usepackage[multiple]{footmisc}
\usepackage{graphicx}
\usepackage{natbib}
\usepackage{color}
\usepackage[dvipsnames]{xcolor}
\usepackage{pdflscape}
\usepackage{booktabs}
\usepackage[anythingbreaks]{breakurl}
\usepackage{setspace}
\usepackage{siunitx}
\usepackage[labelsep=period]{caption}
\usepackage{subcaption}
\usepackage{tabulary}
\usepackage{threeparttablex,makecell}
\usepackage{longtable}
\usepackage{multirow}
\usepackage{enumitem}
\usepackage{multibib}
\newcites{annex}{References}
\usepackage{floatrow}
\floatstyle{plaintop}
\restylefloat{table}
\restylefloat{figure}
\usepackage{authblk}
\usepackage{bbm}

\usepackage{ulem}
\usepackage[hyphens]{url}
\definecolor{myblue}{rgb}{0.15, 0.23, 0.89}
\usepackage{hyperref}
\hypersetup{
	linkcolor  = myblue,
	citecolor  = myblue,
	urlcolor   = myblue,
	colorlinks = true
}
\usepackage{cleveref}
\newcommand{\bi}{\begin{itemize}}
\newcommand{\ei}{\end{itemize}}
\newcommand{\be}{\begin{enumerate}}
\newcommand{\ee}{\end{enumerate}}

\newcolumntype{L}[1]{>{\raggedright\let\newline\\arraybackslash\hspace{0pt}}m{#1}}
\newcolumntype{C}[1]{>{\centering\let\newline\\arraybackslash\hspace{0pt}}m{#1}}
\newcolumntype{R}[1]{>{\raggedleft\let\newline\\arraybackslash\hspace{0pt}}m{#1}}

\newtheorem{asu}{Assumption}

\newcommand\percentage[2][round-precision = 1]{
    \SI[round-mode = places,
        scientific-notation = fixed, fixed-exponent = 0,
        output-decimal-marker={.}, #1]{#2e2}%
}

\makeatletter
\newcommand{\cellref}[2]{
  \def\@currentlabel{\percentage{#1}}\label{#2}%
}
\makeatother

\makeatletter
\newcommand{\cellreff}[2]{
  \def\@currentlabel{#1}\label{#2}%
}
\makeatother

  \makeatletter
  \def\title@font{\Large}
  \let\ltx@maketitle\@maketitle
  \def\@maketitle{\bgroup%
    \let\ltx@title\@title%
    \def\@title{\resizebox{\textwidth}{!}{%
      \mbox{\title@font\ltx@title}%
    }}%
    \ltx@maketitle%
  \egroup}
\makeatother

\newcommand{\sym}[1]{#1}

\begin{document}
\begin{titlepage}
    \title{Can Desegregation Close the Racial Gap in High School Coursework?\footnote{ritika.sethi@rice.edu, 
Department of Economics, Rice University,
Houston, TX 77005. I thank Rossella Calvi, Flavio Cunha, Maura Coughlin, Dibya Mishra, Isabelle Perrigne, Robin Sickles, Xun Tang, and Matthew Thirkettle for their feedback. I extend my appreciation to the participants at the AEA CSWEP, SEA Annual Meeting, AEFP Annual Conference, and RES Annual Conference for their valuable comments. This research uses data from THEOP, a project directed by Marta Tienda, at Princeton University, in collaboration with Teresa A. Sullivan, formerly at the University of Texas at Austin. This research was supported by grants from the Ford, Mellon and Hewlett Foundations and NSF (Grant  SES-0350990). We gratefully acknowledge institutional support from Princeton University's Office of Population Research (NICHD Grant  R24 H0047879). Special acknowledgement is due to Dawn Koffman, for preparing and documenting data for public use, and to Sunny X. Niu who has been a stalwart collaborator and co-investigator. 
}}
    \author[1]{Ritika Sethi}
    \affil[1]{Department of Economics, Rice University}
    \date{\today}
    \maketitle

    \begin{abstract}
        \noindent 
        This paper examines the interplay between desegregation, institutional bias, and individual behavior in education. Using a game-theoretic model that considers race-heterogeneous social incentives, the study investigates the effects of between-school desegregation on within-school disparities in coursework. The analysis incorporates a segregation measure based on entropy and proposes an optimization-based approach to evaluate the impact of student reassignment policies. The results highlight that Black and Hispanic students in predominantly White schools, despite receiving less encouragement to apply to college, exhibit higher enrollment in college-prep coursework due to stronger social incentives from their classmates' coursework decisions.
        \\
        \vspace{0in}
        \bigskip
    \end{abstract}

    \setcounter{page}{0}
    \thispagestyle{empty}
\end{titlepage}

\pagebreak
\newgeometry{right=1in,left=1in,top=1in,bottom=1in}
\setstretch{1.5}
\sloppy
\doublespacing

\section{Introduction}\label{intro}
School desegregation is a topic of immense importance, capturing the attention of policymakers and researchers alike. While desegregation efforts often focus on achieving diversity between schools, examining whether these initiatives lead to equitable outcomes within schools is crucial. Particularly, the under-representation of Black and Hispanic students in advanced math and science courses in high school persists \citep*{civilrights,francis2021separate}. This under-representation holds notable implications for the future educational and employment prospects of these students, potentially contributing to over-representation in remedial math in college, under-representation in STEM majors and careers, and lower college completion rates \citep*{riegle2010questioning,todd2021understanding}.

Extensive research has examined the consequences of desegregation on educational outcomes \citep*{angrist2004does,guryan2004desegregation,hanushek2009new,johnson2011long}. While the existing literature generally suggests that minority students benefit from attending racially integrated schools, the impact of between-school desegregation on course-taking presents a more nuanced and complex picture. A notable observation is the existence of a "segregation paradox," wherein Black and Hispanic students are often underrepresented in advanced courses when enrolled in racially diverse schools. For instance, \citet*{kelly2009black} finds that Black students are more likely to enroll in low-track mathematics when they are in the minority at schools, based on data from the National Education Longitudinal Study. Similarly, \citet*{card2007racial} report a higher racial gap in honors courses in racially-balanced schools using data on SAT test-takers. Another study by \citet*{clotfelter2021school} utilizing administrative data from public schools in North Carolina demonstrates a higher racial gap in advanced math course enrollment in less segregated schools. 

Past literature has proposed factors such as behavioral disengagement due to the social stigma of ``acting white'', ``integration fatigue'', or ``frog pond'' as explanations for why Black and Hispanic students enrolled in White-predominant or racially balanced schools tend to experience worse outcomes \citep*{rothstein2008affirmative,fryer2010empirical,ackert2018segregation}. Another explanation posits that limited access to institutional knowledge on maximizing academic opportunities may lead to less strategic course-taking among Black and Hispanic students \citep*{casey2018academic}. However, a notable exception to this pattern is found in \citeauthor*{billings2014school}'s \citeyearpar{billings2014school} analysis of the impact of race-based busing on educational outcomes in a school district in North Carolina. The study reveals that students, regardless of their race, are less likely to enroll in advanced courses when assigned to Black/Hispanic-majority schools.

In general, investigating the impact of desegregation on within-school representation is a complex task. Desegregation results in changes in student composition within schools. The composition of the student body can impact information-sharing among students, and teachers' perceptions about students' abilities. At the same time, desegregation can include moving costs, changes in school quality, and compensatory institutional responses \citep*{hanushek2009new,johnson2011long}. For example, desegregation may require students to transfer to different schools, resulting in logistical challenges and disruptions to their educational continuity. Moreover, integrating students from different backgrounds may necessitate school policies, resources, and curriculum changes to ensure an inclusive and equitable learning environment. Disentangling these mechanisms can be a challenging endeavor. 

Endogeneity in students' decision-making is another crucial but often overlooked aspect in previous studies. Students' decisions may be influenced by their classmates' choices, leading to strategic decision-making rather than isolated individual choices \citep*{casey2018academic}. Factors such as collaborative study groups and the relative grading system can incentivize students to make decisions strategically. This strategic behavior is particularly relevant in states where college admission outcomes depend on class rank. Our analysis addresses this endogeneity issue by modeling high school students' decisions to take the college-prep coursework as a simultaneous-move game within their graduating class. Based on the available information, each student forms beliefs about their classmates' being encouraged to apply to college and their coursework decisions. As a result, the probabilities of students choosing the college-prep coursework are determined endogenously. We employ the concept of Bayesian Nash equilibrium to characterize these choice probabilities.

Our modeling framework offers an in-depth understanding of the implications of desegregation on coursework decisions. Firstly, it acknowledges the heterogeneity of students' beliefs, recognizing that students respond to individual behavior rather than relying solely on group averages. Secondly, the model accounts for race-specific social incentives between and within racial groups. A model assuming homogeneous social incentives would be overly restrictive in this context. Thirdly, the model incorporates the influence of teacher encouragement on students' coursework decisions, capturing the role of institutional biases in shaping individual behavior. Additionally, the model considers teachers' decisions as functions of the student composition, allowing for direct implications of desegregation on institutional biases. Lastly, the model addresses school-specific unobserved heterogeneity by comparing adjacent graduating classes within the same school, effectively isolating the impact of changes in student composition from inherent differences across schools.

To investigate the impact of desegregation on teachers' and students' decisions, we conduct counterfactual simulations by manipulating the level of segregation, as measured by \citet{theil1972statistical} entropy index. This index quantifies the deviation of the racial composition within specific schools from the overall racial composition in the state. By solving an optimization problem, we seek alternative distributions of student compositions across schools that closely resemble the observed distribution while keeping school capacities, unobserved quality, and overall racial shares constant. Then, we randomly reassign students across schools to construct an alternative distribution of school-level student compositions and hence, an alternative level of segregation between schools. As between-school segregation decreases, we observe a decrease in the probability of teachers encouraging Black and Hispanic students to apply to college. On the other hand, the probability of these students taking college-prep coursework increases. Larger social incentives arising from the coursework decisions of their White classmates drive this increase.

This study makes significant contributions to three strands of literature. Firstly, it advances the literature on modeling educational decision-making, building upon previous studies that have utilized discrete choice models to explore various aspects of education, such as career choices and occupational decisions \citep{keane1997career,arcidiacono2004ability,de2019dynamic}. This study takes a step further by incorporating strategic coursework selections made by high school students into the analysis. Secondly, the paper contributes to the relatively limited body of research investigating desegregation's effects on educational decision-making \citep*{card2007racial,mookherjee2010aspirations,clotfelter2021school,francis2021separate}. This research makes a valuable contribution to a broader literature on the role of social interactions in shaping educational decisions \citep*{zanella2007discrete,de2010identification,tincani2018heterogeneous,wustudent2021}. Understanding how social incentives shape students' decision-making is essential for developing targeted and effective policy interventions. Lastly, this study aligns with the growing body of research that employs game-theoretic approaches to estimate social incentives in various contexts \citep*{lee2014binary,ciliberto2016playing,yang2017social,guerra2020multinomial,lin2021uncovering}. By allowing for heterogeneous social incentives and utilizing an optimization-based method, this research explores the implications of group redistribution on individual behavior in a simultaneous-move game setting. 

The remainder of the paper is structured as follows:  Section \ref{data} provides an overview of the data and institutional background. Then, Section \ref{model} presents the modeling framework used in the analysis. The identification strategy is discussed in Section \ref{identification}, followed by the estimation method in Section \ref{estimation}. The results obtained from the model are presented in Section \ref{results}. Section \ref{counterfactual} explores the outcomes of counterfactual policy experiments. Finally, the paper concludes with Section \ref{conclusion}.

\section{Background and Data}\label{data}

Our study focuses on exploring the coursework decisions of high school students in Texas. We utilize student data from the Texas Higher Education Opportunity Project (THEOP), a research initiative managed by the Office of Population Research at Princeton University \citep*{theop}. THEOP administered surveys to Sophomores and Seniors from a state-representative sample of public schools in Texas in 2002, capturing information on various aspects of their academic journey. The THEOP data has been employed in various recent studies studying college outcomes, including those conducted by \citet*{kapor2020distributional}, \citet*{li2020estimating}, and \citet*{akhtari2020affirmative}.

Students had the option to choose among three high school graduation plans:  the Minimum High School Program (MHSP), Recommended High School Program (RHSP), and Distinguished Achievement Program (DAP) \citep*{texreg}. These plans differ in course requirements to be eligible to graduate high school and serve as pathways to college admission. The MHSP offers the most flexibility in course selection, with only Algebra I, which is the first course in the usual sequence of math coursework, compulsory to fulfill the math coursework requirements. On the other hand, the RHSP additionally requires Geometry and Algebra II, with the option to take more rigorous math courses like Calculus. The DAP has even stricter requirements compared to the RHSP. For example, while less-rigorous courses like ``math models" or ``integrated physics and chemistry" meet the requirements of the RHSP. Such options are not available under the DAP. Generally, the RHSP and the DAP have greater demand in terms of the number and rigor of courses, and completion of these graduation plans fulfills core curriculum requirements for admission to major four-year colleges. Our study defines a student as deciding to take the college-prep coursework if they choose either the RHSP or the DAP as their high school graduation plan.

\begin{table}[!t]\caption{Students' Characteristics and their Coursework Decision}\centering\label{tab:sumstat1}\fontsize{8}{8}\selectfont
\begin{threeparttable}
\begin{tabulary}{\linewidth}{lccccc@{}}
\toprule
 &  (1) & (2) & (3) & (4) & (5) \\ \addlinespace
 & \multicolumn{2}{c}{}  & \multicolumn{3}{c}{College-prep coursework}   \\ 
\cmidrule(lr){2-3} \cmidrule(lr){4-6} 
 &  Mean & S.D. & No & Yes & Difference \\ \midrule
\addlinespace \multicolumn {6}{l}{\textbf{Panel A: Demographic information}} \\ 
 Female  & 0.432\cellref{0.432}{female} & 0.495 & 0.593\cellref{0.593}{female0}  & 0.641\cellref{0.641}{female1}  & $0.049\cellref{0.049}{femaled}^{***}$  \\ 
 White  & 0.539\cellref{0.539}{white} & 0.498 & 0.597\cellref{0.597}{white0}  & 0.628\cellref{0.628}{white1}  & $0.031\cellref{0.031}{whited}^{***}$  \\ 
 Mother attended college  & 0.386\cellref{0.386}{momschool} & 0.487 & 0.545\cellref{0.545}{momschool0}  & 0.723\cellref{0.723}{momschool1}  & $0.179\cellref{0.179}{momschoold}^{***}$  \\ 
 Family owns home  & 0.648\cellref{0.648}{homeowner} & 0.478 & 0.527\cellref{0.527}{homeowner0}  & 0.661\cellref{0.661}{homeowner1}  & $0.133\cellref{0.133}{homeownerd}^{***}$  \\ 
 Two parents household  & 0.492\cellref{0.492}{bothparent} & 0.500 & 0.562\cellref{0.562}{bothparent0}  & 0.667\cellref{0.667}{bothparent1}  & $0.104\cellref{0.104}{bothparentd}^{***}$  \\ 
\addlinespace \multicolumn {6}{l}{\textbf{Panel B: Course Grades}} \\ 
 Got an A in English  & 0.300\cellref{0.300}{enggrade} & 0.458 & 0.551\cellref{0.551}{enggrade0}  & 0.760\cellref{0.760}{enggrade1}  & $0.208\cellref{0.208}{enggraded}^{***}$  \\ 
 Got an A in Math  & 0.223\cellref{0.223}{mathgrade} & 0.416 & 0.575\cellref{0.575}{mathgrade0}  & 0.748\cellref{0.748}{mathgrade1}  & $0.173\cellref{0.173}{mathgraded}^{***}$  \\ 
 Got an A in History  & 0.349\cellref{0.349}{historygrade} & 0.477 & 0.535\cellref{0.535}{historygrade0}  & 0.759\cellref{0.759}{historygrade1}  & $0.224\cellref{0.224}{historygraded}^{***}$  \\ 
 Got an A in Science  & 0.270\cellref{0.270}{sciencegrade} & 0.444 & 0.561\cellref{0.561}{sciencegrade0}  & 0.756\cellref{0.756}{sciencegrade1}  & $0.196\cellref{0.196}{sciencegraded}^{***}$  \\ 
 Class rank: top thirty decile  & 0.366\cellref{0.366}{topthirty} & 0.482 & 0.504\cellref{0.504}{topthirty0}  & 0.804\cellref{0.804}{topthirty1}  & $0.300\cellref{0.300}{topthirtyd}^{***}$  \\ 
\addlinespace \multicolumn {6}{l}{\textbf{Panel C: Discussions with Counsellor}} \\ 
 Discussed courses  & 0.737\cellref{0.737}{_q26a} & 0.440 & 0.517\cellref{0.517}{_q26a0}  & 0.648\cellref{0.648}{_q26a1}  & $0.131\cellref{0.131}{_q26ad}^{***}$  \\ 
 Discussed future education  & 0.440\cellref{0.440}{_q26e} & 0.496 & 0.571\cellref{0.571}{_q26e0}  & 0.668\cellref{0.668}{_q26e1}  & $0.097\cellref{0.097}{_q26ed}^{***}$  \\ 
 Discussed personal problems  & 0.152\cellref{0.152}{_q26b} & 0.359 & 0.622\cellref{0.622}{_q26b0}  & 0.567\cellref{0.567}{_q26b1}  & $-0.055\cellref{0.055}{_q26bd}^{***}$  \\ 
 Discussed discipline problems  & 0.120\cellref{0.120}{_q26c} & 0.325 & 0.629\cellref{0.629}{_q26c0}  & 0.503\cellref{0.503}{_q26c1}  & $-0.126\cellref{0.126}{_q26cd}^{***}$  \\ 
 Discussed jobs  & 0.159\cellref{0.159}{_q26d} & 0.366 & 0.615\cellref{0.615}{_q26d0}  & 0.605\cellref{0.605}{_q26d1}  & $-0.011\cellref{0.011}{_q26dd}^{}$  \\ 
\addlinespace \multicolumn {6}{l}{\textbf{Panel D: Encouragement from Teacher}} \\ 
 Encouraged for college  & 0.738\cellref{0.738}{_q28a} & 0.439 & 0.472\cellref{0.472}{_q28a0}  & 0.664\cellref{0.664}{_q28a1}  & $0.191\cellref{0.191}{_q28ad}^{***}$  \\ 
 Encouraged for vocational school  & 0.208\cellref{0.208}{_q28b} & 0.406 & 0.621\cellref{0.621}{_q28b0}  & 0.587\cellref{0.587}{_q28b1}  & $-0.034\cellref{0.034}{_q28bd}^{***}$  \\ 
 Encouraged for apprenticeship  & 0.218\cellref{0.218}{_q28c} & 0.413 & 0.616\cellref{0.616}{_q28c0}  & 0.607\cellref{0.607}{_q28c1}  & $-0.009\cellref{0.009}{_q28cd}^{}$  \\ 
 Encouraged for military service  & 0.195\cellref{0.195}{_q28d} & 0.397 & 0.622\cellref{0.622}{_q28d0}  & 0.579\cellref{0.579}{_q28d1}  & $-0.043\cellref{0.043}{_q28dd}^{***}$  \\ 
 Encouraged for jobs after high school  & 0.312\cellref{0.312}{_q28e} & 0.463 & 0.623\cellref{0.623}{_q28e0}  & 0.594\cellref{0.594}{_q28e1}  & $-0.029\cellref{0.029}{_q28ed}^{***}$  \\ 
\addlinespace \multicolumn {6}{l}{\textbf{Panel E: What matters for College Admissions}} \\ 
 Course grades matter  & 0.676\cellref{0.676}{_q32a} & 0.468 & 0.542\cellref{0.542}{_q32a0}  & 0.648\cellref{0.648}{_q32a1}  & $0.106\cellref{0.106}{_q32ad}^{***}$  \\ 
 Coursework matter  & 0.507\cellref{0.507}{_q32b} & 0.500 & 0.562\cellref{0.562}{_q32b0}  & 0.663\cellref{0.663}{_q32b1}  & $0.101\cellref{0.101}{_q32bd}^{***}$  \\ 
 Class rank matter  & 0.458\cellref{0.458}{_q32e} & 0.498 & 0.567\cellref{0.567}{_q32e0}  & 0.669\cellref{0.669}{_q32e1}  & $0.102\cellref{0.102}{_q32ed}^{***}$  \\ 
 HS diploma matter  & 0.755\cellref{0.755}{_q32g} & 0.430 & 0.513\cellref{0.513}{_q32g0}  & 0.646\cellref{0.646}{_q32g1}  & $0.133\cellref{0.133}{_q32gd}^{***}$  \\ 
 Race matters  & 0.098\cellref{0.098}{_q32i} & 0.298 & 0.619\cellref{0.619}{_q32i0}  & 0.568\cellref{0.568}{_q32i1}  & $-0.050\cellref{0.050}{_q32id}^{***}$  \\ 
\addlinespace \multicolumn {6}{l}{\textbf{Panel F: Friends characteristics}} \\ 
 Do well in school  & 0.670\cellref{0.670}{_q22a} & 0.470 & 0.502\cellref{0.502}{_q22a0}  & 0.669\cellref{0.669}{_q22a1}  & $0.167\cellref{0.167}{_q22ad}^{***}$  \\ 
 Plan to go to college  & 0.702\cellref{0.702}{_q22b} & 0.457 & 0.443\cellref{0.443}{_q22b0}  & 0.686\cellref{0.686}{_q22b1}  & $0.242\cellref{0.242}{_q22bd}^{***}$  \\ 
 Think it's important to work hard  & 0.459\cellref{0.459}{_q22c} & 0.498 & 0.553\cellref{0.553}{_q22c0}  & 0.685\cellref{0.685}{_q22c1}  & $0.132\cellref{0.132}{_q22cd}^{***}$  \\ 
 Participate in extra-curriculars  & 0.534\cellref{0.534}{_q22d} & 0.499 & 0.505\cellref{0.505}{_q22d0}  & 0.708\cellref{0.708}{_q22d1}  & $0.203\cellref{0.203}{_q22dd}^{***}$  \\ 
 \midrule Obs  &  \multicolumn{2}{c}{28,579} &  & \\ 

\\
\bottomrule
\end{tabulary}
\begin{tablenotes}[flushleft, online, normal, para]\scriptsize
\emph{Note}: This table provides student-level characteristics in Columns (1) and (2) as means and standard deviations. Each variable represents a binary indicator for a student characteristic, such as Female, for whether the student's gender is female. Columns (3) and (4), respectively, show group averages for those who take the college-prep coursework and those who do not, and Column (5) presents the t-test results. For instance, the first row displays the average college-prep coursework takeup for males in Column (4) and for females in Column (5).
Significance levels: $^{}p<$0.1; $^{}p<$0.05; $^{}p<$0.01.\\
\emph{Data source}: Texas Higher Education Opportunity Project (THEOP), Wave I, 2002.
 \end{tablenotes}
\end{threeparttable}
\end{table}

The THEOP survey gathers data on students' demographics, course enrollment, grades, class rank, high school graduation plan choices, and college aspirations. Furthermore, the survey includes questions about students' interactions with teachers, specifically regarding academic matters like whether the teacher encourages them to apply to college. This provides a distinct opportunity to examine students' educational decision-making within the same class and investigate the influence of teachers on these decisions. Table \ref{tab:sumstat1} highlights the relationship between student characteristics and the decision to take the college-prep coursework. In Panel A, female and White students with college-educated mothers, from home-owner families, and living in two-parent households are more likely to select college-prep coursework.

In Panel B of  Table \ref{tab:sumstat1}, higher course grades and class rank in the top thirty percentile correlate positively with college-prep coursework selection, which aligns with the benefits offered in Texas, such as automatic admission to in-state public universities for students in the top ten percentile and financial aid eligibility for those in the top twenty-five percentile.

In Panel C, conversations about courses and future education plan positively correlate with the decision to take the college-prep coursework. At the same time, discussions regarding personal or disciplinary problems show a negative correlation. Panel D examines the relationship between teacher encouragement and coursework decisions. Students encouraged to pursue college are likelier to opt for college-prep coursework, while encouragement for other post-high school options (such as vocational school) shows a negative association.

The THEOP survey asked students to report how much importance they attribute to different factors in the college admissions process. In Panel E, higher importance ratings for course grades, coursework, class rank, and a high school diploma positively correlate with choosing college-prep coursework. Students' social circles also influence coursework decisions. In Panel F, having three or more friends who excel academically, plan to attend college, value hard work, and engage in extracurricular activities positively correlates with the probability of choosing college-prep coursework.

These findings highlight the impact of factors such as family background, academic performance, teacher interactions, institutional knowledge, and social influences on students' coursework decision-making. Table \ref{tab:logit} illustrates how these factors vary with the racial composition of the class. Controlling for gender, course grades, and mother's education, and accounting for school and graduating class fixed effects, the analysis shows that as the share of White students increases, the probability of teachers encouraging Black and Hispanic students to apply to college, the probability of Black and Hispanic students having friends with college plans, and their probability of taking the college-prep coursework decrease. In contrast, the probability of Hispanic students perceiving race as a factor in college admissions increases.

Changes in student composition may not only impact teachers' perceptions of students' abilities, leading to variations in teacher recommendations but may also directly influence students' social incentives in selecting the college-prep coursework, including their tendency to coordinate with or deviate from their classmates' behavior. To exhaustively explore the implications of changes in student composition, we develop a model that integrates both students' coursework and teachers' encouragement decisions in the subsequent section.

\begin{table}[!t]\caption{Racial Composition,  Perceptions,  Aspirations, and Coursework Decision}\centering\label{tab:logit}\fontsize{8}{8}\selectfont
\begin{threeparttable}
\begin{tabulary}{\linewidth}{lcccc@{}}
\toprule
& \makecell{Teacher encourages\\for college} & \makecell{Friends have\\college plans} & \makecell{Race matters \\ for admissions} & \makecell{College-prep \\ coursework} \\ \midrule
Share of White students&      -0.433         &       0.258         &      -1.018\sym{*}  &       0.785\sym{**} \\
                    &     (0.371)         &     (0.340)         &     (0.532)         &     (0.344)         \\
\addlinespace
Black $\times$ Share of White students&      -0.551\sym{**} &      -1.202\sym{***}&      -0.367         &      -0.480\sym{**} \\
                    &     (0.218)         &     (0.219)         &     (0.309)         &     (0.203)         \\
\addlinespace
Hispanic $\times$ Share of White students&      -0.721\sym{***}&      -1.452\sym{***}&       0.466\sym{**} &      -0.914\sym{***}\\
                    &     (0.154)         &     (0.144)         &     (0.223)         &     (0.141)         \\
\addlinespace
Black               &       0.423\sym{***}&       0.874\sym{***}&       0.689\sym{***}&       0.344\sym{***}\\
                    &     (0.134)         &     (0.128)         &     (0.187)         &     (0.120)         \\
\addlinespace
Hispanic            &       0.381\sym{***}&       0.696\sym{***}&       0.139         &       0.461\sym{***}\\
                    &     (0.088)         &     (0.078)         &     (0.128)         &     (0.078)         \\
\addlinespace
Female              &       0.168\sym{***}&       0.353\sym{***}&      -0.090\sym{**} &       0.116\sym{***}\\
                    &     (0.027)         &     (0.028)         &     (0.042)         &     (0.026)         \\
\addlinespace
Course grade: A     &       0.415\sym{***}&       0.478\sym{***}&      -0.086\sym{**} &       0.770\sym{***}\\
                    &     (0.027)         &     (0.030)         &     (0.043)         &     (0.028)         \\
\addlinespace
Mother attended college&       0.556\sym{***}&       0.757\sym{***}&       0.097\sym{**} &       0.702\sym{***}\\
                    &     (0.028)         &     (0.032)         &     (0.044)         &     (0.029)         \\
\midrule Mean & 0.309 & 0.702 & 0.098 & 0.614  \\
Obs & 28,579 & 28,579 & 28,579 & 28,579  \\

\\
\bottomrule
\end{tabulary}
\begin{tablenotes}[flushleft, online, normal, para]\scriptsize
\emph{Note}: This table presents logit regressions for the probabilities of teacher encouragement (Column 1), student's friends aspiring to go to college (Column 2), student's perception of race in college admissions (Column 3), and student's choice of college-prep coursework (Column 4). School and graduating class fixed effects are included.
Significance levels: $^{}p<$0.1; $^{}p<$0.05; $^{}p<$0.01.\\
\emph{Data source}: Texas Higher Education Opportunity Project (THEOP), Wave I, 2002.
 \end{tablenotes}
\end{threeparttable}
\end{table}

\section{Model}\label{model}

\subsection{Student's decision}
In the context of a high school, $s \in \{1,\ldots, S\}$, and multiple graduating classes, $g \in \{1,\ldots, G\}$, within each high school, student $i$ faces the decision whether to take the college-prep coursework, $a_{igs}\in\{0,1\}$. Normalizing the utility of not taking the college-prep coursework to zero, student $i$ in graduating class $g$ in high $s$ decides to take the college-prep coursework if:
\begin{equation}\label{eq:studentdecision}
x_{igs} \beta + b_{igs} \alpha  + \kappa_g + \gamma_s + \frac{1}{N_{gs}-1}\sum_{j\in \mathcal{N}_{gs}\setminus \{i\}} \lambda_{ij}
\mathbb{E}\left(a_{jgs}\mid x_{gs} , b_{igs}, \epsilon_{igs}\right)  +\epsilon_{igs} > 0.
\end{equation}
In this equation, $\mathcal{N}_{gs} \equiv \{1,\ldots,N_{gs}\}$ represents the set of students in graduating class $g$ in high school $s$. Several factors influence this decision. Firstly, the student's characteristics $x_{igs}\in \mathcal{X}$, including race, gender, academic performance, and parental education, which are known to everyone in the same graduating class and high school. Secondly, whether the student is encouraged by their teacher to apply to college, denoted as $b_{igs}$, which is only known to the student (and observed in the data,) but not to their classmates. Thirdly, an unobserved component, $\epsilon_{igs}$, influences the student's motivation or hindrance to taking the college-prep coursework. This component is known only to the student and not to their classmates. Fourthly, the student's beliefs about their classmates' coursework decisions, denoted as $\mathbb{E}\left(a_{jgs} \mid x_{gs}, b_{igs},\epsilon_{igs}\right)$, shape the student's coursework decisions. It is important to note that this belief is conditional on the information available to the student, ($x_{gs}, b_{igs},\epsilon_{igs}$), where $x_{gs}$ represents the characteristics of all students in the class. Finally, the student's decision is further influenced by unobserved graduating class-specific and school-specific factors, captured by $\kappa\equiv \{\kappa_1,\ldots,\kappa_G\}$ and $\gamma \equiv \{\gamma_1,\ldots,\gamma_S\}$, respectively.

The term $\sum_{j\in \mathcal{N}_{gs}\setminus \{i\}} \lambda_{ij} \mathbb{E}\left(a_{jgs}\mid x_{gs} , b_{igs}, \epsilon_{igs}\right)$ denotes the weighted average of the expected decisions of student $i$'s classmates. The parameter $\lambda_{ij}$ captures the heterogeneity in the utility associated with beliefs about classmates' decisions, considering student $i$'s race and their classmates' race. Specifically, if $x^r_{igs}$ represents student $i$'s race, then $\lambda_{ij} = \lambda_{r_1 r_2} \text{ if } x^r_{igs} = r_1 \text{ and } x^r_{igs} = r_2, \text{ where } r_1,r_2 \in \{\text{White}, \text{Black}, \text{Hispanic}\}.$ A positive value of $\lambda_{ij}$ indicates that student $i$ prefers to coordinate with student $j$'s coursework decision, while a negative value indicates that student $i$ prefers to deviate from student $j$'s coursework decision.

\subsection{Teacher's decision}
Students do not have direct knowledge of whether their classmates are encouraged by the teacher to apply to college ($b_{igs}$). Normalizing the utility of not encouraging a student to apply to college, the teacher encourages student $i$ in graduating class $g$ in high school $s$ to apply to college if:

\begin{equation}\label{eq:teacherdecision}
x_{igs}\delta + \xi_g + \zeta_s +  \frac{1}{N_{gs}-1} \sum_{j\in \mathcal{N}_{gs}\setminus \{i\}} \rho_{ij} + \eta_{igs} > 0,
\end{equation}
where $\eta_{igs}$ is a student-specific unobserved factor that affects the teacher's decision in addition to the student's observed characteristics $x_{igs}$. The decision also depends on graduating class-specific and school-specific unobserved factors, captured by $\xi \equiv \{\xi_1,\ldots,\xi_G\}$ and $\zeta \equiv \{\zeta_1,\ldots,\zeta_S\}$, respectively.  Additionally, the teacher takes into account the racial composition of the class; $\rho_{ij}$ allows for heterogeneity in the teacher's decisions based on the student's race and their classmate's race: $\rho_{ij} = \rho_{r_1 r_2} \text{ if } x^r_{igs} = r_1 \text{ and } x^r_{igs} = r_2, \text{ where } r_1,r_2 \in \{\text{White}, \text{Black}, \text{Hispanic}\}.$ A positive value of $\rho_{ij}$ indicates that student $i$ is more likely to be encouraged to apply to college as the share of students of the same race as student $j$ increases.

\subsection{Equilibrium} 
Consider student $k$ in graduating class $g$ in high school $s$. We want to determine the probability of student $i$ taking college-prep coursework from the perspective of student $k$:

\begin{equation}\label{eq:studenteqm}
\resizebox{.9\hsize}{!}{$
\begin{aligned}
 \Pr \left(a_{i  g s} = 1\mid x_{gs},b_{gs},\epsilon_{k g s}, \eta_{kgs}\right) 
& =  
\mathbb{E} ( x_{igs} \beta + b_{igs} \alpha   + \kappa_g +  \gamma_s 
\\ & + 
 \frac{1}{N_{gs}-1}\sum_{j\in \mathcal{N}_{gs}\setminus \{i\}} \lambda_{ij}
\Pr\left(a_{jgs} = 1\mid x_{gs} , b_{igs} , \epsilon_{igs} , \eta_{kgs}\right)  +\epsilon_{igs}>0 \}\mid x_{gs}, b_{kgs},\epsilon_{k g s},\eta_{kgs} ).
\end{aligned}
$}
\end{equation}
We assume that $\epsilon_{igs}$ and $\eta_{igs}$ are independent and identically distributed within and across graduating classes and schools and  independent from $x_{gs}$. This assumption implies that the information available privately to student $k$ , i.e., $b_{kgs}, \epsilon_{kgs}, \eta_{kgs}$, does not provide any additional insight into student $i$'s coursework decision or the teacher's decision to encourage student $i$ to apply to college. In other words, student $k$ forms beliefs about student $i$'s getting encouraged to apply to college and choosing the college-prep coursework based solely on the publicly available information $x_{gs}$. Now, the probability of student $i$ taking college-prep coursework based on the information available to student $k$ is
\begin{equation}\label{eq:bestresponse}
\begin{aligned}
\Pr \left(a_{i  g s} = 1 \mid x_{gs} \right) 
& =  
\sum_{b \in \{0,1\}}\Pr(b_{igs} = b \mid x_{gs})\mathbb{E} ( x_{igs} \beta + b\alpha   + \kappa_g +  \gamma_s 
\\ & + 
 \frac{1}{N_{gs}-1}\sum_{j\in \mathcal{N}_{gs}\setminus \{i\}} \lambda_{ij}
\mathbb{E}\left(a_{jgs}\mid x_{gs} \right)  +\epsilon_{igs}>0 \}\mid x_{gs}  ).
\end{aligned}
\end{equation}
In other words, $\Pr \left(a_{i  g s} = 1\mid x_{gs} \right)$ is the probability that student $i$ decides to take the college-prep coursework from the perspective of their classmates. In equilibrium, this is student $i$'s best response to their beliefs regarding the decisions of other students in their class. Let $\sigma_{igs}$ represent the probability of student $i$ taking the college-prep coursework. A pure-strategy Bayesian-Nash equilibrium of this simultaneous-move game of incomplete information within graduating class $g$ in high school $s$ is defined by a vector of conditional choice probabilities denoted as $\sigma_{gs}\equiv\{\sigma_{1gs},\ldots,\sigma_{N_{gs} gs}\}:\mathcal{X}  \rightarrow \{0,1\}^{N_{gs}}$. 

The subsequent discussion will establish conditions that guarantee the uniqueness of the equilibrium. These conditions ensure that the best response function, represented by Equation (\ref{eq:bestresponse}), exhibits properties of a contraction mapping. By examining the estimation results, we will determine whether we can rule out the possibility of multiple equilibria while refraining from imposing constraints on the parameter space.

\section{Identification}\label{identification}
In this section, we discuss the strategy to identify the parameters associated with the teacher's decisions regarding encouraging students to apply to college, $\theta_1\equiv (\delta,\xi,\zeta,\rho)$ and students' decisions regarding taking the college-prep coursework, $\theta_2 \equiv (\beta,\alpha,\kappa,\gamma,\lambda)$. 

The conditional probabilities of a teacher encouraging the student to apply to college and of a student taking the college-prep coursework are identified by examining the sample moments from the data. The parameters $\gamma$ and $\zeta$ are associated with school-specific unobserved heterogeneity.\footnote{We normalize $\gamma_1 =0$ and $\zeta_1 = 0$. To ensure the consistency of the estimates, it is necessary to have a sufficiently large number of students in each school \citep*{lee2014binary}. Therefore, the analysis is based on high schools with at least 100 students.} They are identified by studying variations within schools, as the sample encompasses multiple independent schools. The parameters $\xi$ and $\kappa$ are associated with graduating class-specific unobserved heterogeneity.\footnote{Each high school has two graduating classes in the sample, the Sophomore cohort and the Senior cohort. We normalize $\xi_1 = 0$ and $\zeta_1 = 0$.} They are identified by studying variations within graduating classes. The parameters $\beta$ and $\delta$ are associated with the student-specific observed characteristics. They are identified by analyzing variations among students. The social utility parameters $\rho$'s and $\lambda$'s are identified by studying schools with students from different racial backgrounds.\footnote{We normalize $\rho_{WW}=\rho_{BW}=\rho_{HW}=0$.} Within-race social utility parameters are determined by examining schools with multiple students of the same race. Conversely, between-race social utility parameters are identified by studying schools with students from different racial backgrounds.

Next, we outline the assumptions that underpin our analysis.
\begin{asu}\label{as:1}
We assume that $\epsilon_{igs}$ and $\eta_{igs}$ are independent and identically distributed within and across graduating classes and high schools, and independent of $x_{gs}$. They are also assumed to be drawn from a type I extreme value distribution with a location parameter of 0 and a scale parameter of 1.
\end{asu}
This assumption implies that a student's private information does not provide additional insight into the teacher's decision regarding encouraging another student or another student's coursework decision. In other words, the variations in the student composition across graduating classes within a school can be treated as random. We can verify how reasonable this assumption is by leveraging the availability of data on adjacent graduating classes within the same school. Including school and graduating year fixed effects in the utility specification allows us to compare students with similar characteristics who are close in age and attending the same school while capturing variations in neighborhood and school factors.

 \begin{figure}[!t]\caption{Distribution of Racial Composition within Schools}\label{fig:rcvar2}\centering 
  \begin{minipage}{0.5\linewidth}
\includegraphics[width=\textwidth]{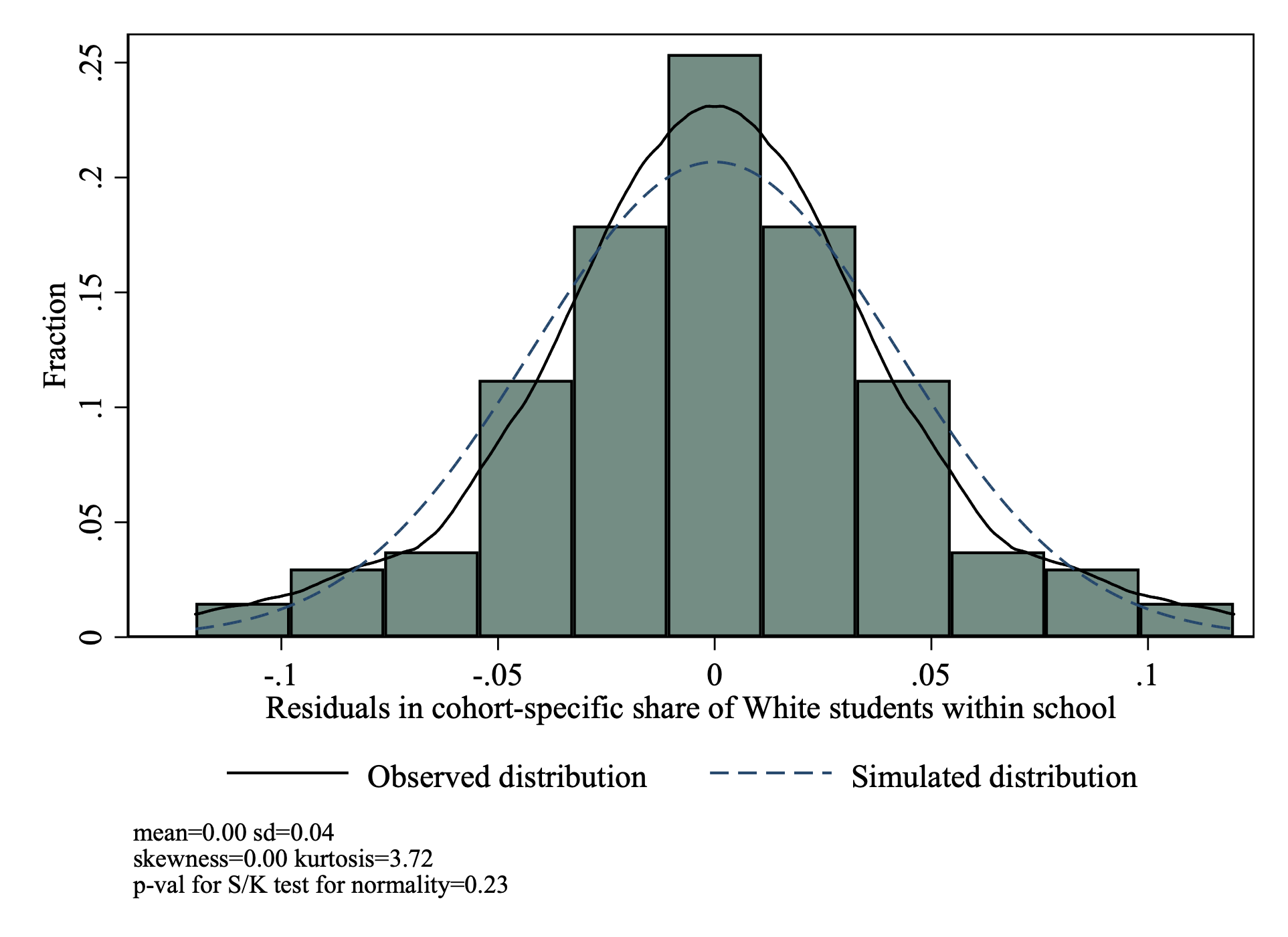}
\end{minipage}
 \begin{minipage}{0.5\linewidth}\scriptsize
\emph{Note:} 
This figure shows the distribution of the variation in the proportions of White students between the cohorts within a school. This figure plots the distribution of residualized racial shares, and includes a simulated normal distribution with the same standard deviation. 
 \\
\emph{Data source}: Texas Higher Education Opportunity Project (THEOP), Wave I, 2002
 \end{minipage}
\end{figure}

As depicted in Figure \ref{fig:rcvar2}, the variation in racial composition after controlling for school fixed and graduating class effects resembles a normal distribution. The observed distribution closely resembles the simulated normal distribution. This approach, which assumes random assignment conditional on observable student factors and school fixed effects, has been employed in various studies \citep{hoxby2000peer,lee2014binary,anelli2019effects,caetano2017school,hoekstra2018peer,cools2019girls,borgen2022counteracting}.

\begin{asu}\label{as:2}
We assume that $\frac{1}{N_{gs}-1}\sum_{j \in \mathcal{N}_{gs} \setminus \{1\}} \mid \lambda_{ij} \mid \leq 4 ; \forall ; g \in \{1,\ldots,G\} , s \in \{1,\ldots, S\}$.
\end{asu}
This assumption constrains the strength of social utilities in the model, ensuring that the best response function in Equation (\ref{eq:bestresponse}) is a contraction mapping. This a common tool employed to establish the uniqueness of equilibrium in Bayesian games \citep{lee2014binary, lin2021uncovering}. This condition guarantees the existence of a unique equilibrium. It is derived by considering the gradient of the vector of choice probabilities and its maximum row sum. It is important to note that we do not impose this condition on the parameter space. We verify whether this condition holds post the estimation of the model.

We can consider the vector of choice probabilities in a generic graduating class and high school, omitting the subscripts $g$ and $s$: $\sigma = (\sigma_1,\ldots,\sigma_{N})$. The gradient for $\sigma $ is given by 

\begin{equation}\label{eq:sigmagradient}
\frac{\partial \sigma}{\partial \sigma^{\prime}}=  
\begin{pmatrix}
0&\frac{1}{N-1}\lambda_{12} \sigma_1(1-\sigma_1) & \dots &\frac{1}{N-1}\lambda_{1N}  \sigma_1(1-\sigma_1))
\\ \vdots & \ddots & \vdots 
\\ \frac{1}{N-1} \lambda_{N1}\sigma_N(1-\sigma_N) & \frac{1}{N-1} \lambda_{N2}\sigma_N(1-\sigma_N)  & \dots & 0
\end{pmatrix}
\end{equation}

Let  $\left\| . \right\|_{\infty}$ denote the maximum row sum of a square matrix. A sufficient condition for equation (\ref{eq:bestresponse}) to be a contraction mapping is that $\left\|\frac{\partial \sigma}{\partial \sigma^{\prime}}\right\|_{\infty} <1$. Note that under the Assumption (\ref{as:1}) that $\epsilon$'s are drawn from a type I extreme value distribution, $\sigma_i(1-\sigma_i)$ has a maximum value of $\frac{1}{4}$. Therefore, it follows that $\sigma$ is a contraction mapping with respect to the $\|.\|_{\infty}$ norm if $\max _{i} \frac{1}{N-1}\sum_{j \neq i } \mid \lambda_{ij}\mid\leq 4.$ 

\begin{asu}\label{as:3}  We assume that $E\left[\operatorname{Var} \left(x_{igs}, \quad b_{igs} , \quad   \frac{1}{N_{gs}-1}\sum\limits_{j \in \mathcal{N}_{gs}\setminus \{i\} }  \mathbb{E}(a_{j gs}\mid x_{gs} ) \right) \right]$ is non-singular. 
\end{asu}
This assumption is a standard rank condition on the equilibrium choice probabilities. It holds because, in general, $\mathbb{E}(a_{jgs}\mid x_{gs} )$ is non-linearly related to the observed characteristics, as it solves a non-linear fixed-point equation in Equation (\ref{eq:bestresponse}) \citep*{yang2017social,lin2021uncovering}.

\section{Estimation}\label{estimation} 
In this section, we outline the estimation procedure for the parameters associated with the teacher's decision, $\theta_1 \equiv (\delta,\xi,\zeta,\rho)$, and the student's decision, $\theta_2\equiv (\beta,\alpha,\kappa,\gamma,\lambda)$. The estimation approach involves a two-step maximum likelihood estimation.

The utility specification includes vector $x_{igs}$ that consist of binary variables representing various characteristics of student $i$. These characteristics include the student's race (White, Black, or Hispanic), gender, whether they are in the top thirty percentile of their class, and whether their mother attended college. These $K (=5)$ binary variables define the state space $\mathcal{X} \equiv \{0,1\}^K$, representing all possible combinations of these characteristics. As a result, there are $2^K$ observed student types. The student's type $t \in \{0,1\}^K$ allows for mapping of the best response function in the state space $2^K$, increasing the computational efficiency considerably.

\subsection{First step: Teacher's decision}
The log-likelihood of student $i$ in graduating class $g$ of high school $s$ getting encouraged to apply to college is:
\begin{equation}\label{eq:teacherlike}
L_{itgs}\left(\theta_1\right) = \mathbbm{1}\left\{b_{itgs}=1\right\}\log\left(\phi_{itgs}\left(x_{gs} ; \theta_1\right)\right) +  \left(1-\mathbbm{1}\left\{b_{itgs}=1\right\}\right)\log\left(1-\phi_{itgs}\left(\theta_1\right)\right)
\end{equation}
where $b_{itgs}$ represents the teacher's (observed) decision whether to encourage student $i$ of type $t$ for college and $\phi_{itgs}\left(x_{gs};\theta_1 \right)$ represents the probability of student $i$ getting encouraged to apply to college. Under the assumption that $\eta_{igs}$ follow a type I extreme value distribution with a location parameter of 0 and a scale parameter of 1, this probability is:
\begin{equation}\label{eq:teacherccp}
\phi_{itgs} \left( x_{gs} ; \theta_1 \right)= \frac{\exp\left(  x_{tgs} \delta + \xi_g + \zeta_s +  \sum_{t^\prime} \rho_{tt^\prime}  w_{tt^\prime gs }  
   \right)}{1+\exp\left(  x_{tgs} \delta + \xi_g + \zeta_s +  \sum_{t^\prime} \rho_{tt^\prime}  w_{tt^\prime gs }    \right)},
\end{equation}
where $w_{tt^\prime gs}=\frac{N_{t^\prime gs} - \mathbbm{1}{t^\prime=t}}{N_{gs}-1}$ represents the share of classmates of type $t^\prime$ from the perspective of a student of type $t$ in graduating class $g$ of high school $s$. We can think of $w_{tt^\prime g s }$ as a term in a matrix representing the social network of graduating class $g$ and high school $s$. To illustrate this, we can define a $N_{tgs}\times N_{tgs}$ matrix, denoted as $w^*_{tt^\prime gs}$, with elements $w^*_{tt^\prime gs}=N_{t^\prime gs}$ if $t^\prime\neq t$ and $w^*_{tt^\prime g}=n_{t^\prime gs}-1$ if $t^\prime=t$. This matrix represents the social network within graduating class $g$, where each student of type $t$ knows $N_{t g}-1$ classmates of the same type (excluding themselves) and $N_{t^\prime g}$ students of type $t^\prime$. By row-normalizing this matrix, we obtain the values of $w_{t^\prime tg}$. In this context, a student of type $t$ assigns a weight of $\frac{N_{tgs} - 1}{N_gs-1}$ to their same-type classmates and $\frac{N_{t^\prime gs}}{N_gs-1}$ to classmates of type $t^\prime$. These weights change when a policy alters the student composition within a graduating class.

Due to the assumption that the unobserved random components $\eta$'s are independent and identically distributed within each graduating class, the likelihoods of individual outcomes are independent of each other, conditional on the observed characteristics. This assumption allows for the summation of likelihoods across students within each graduating class. We estimate the parameters $\theta_1$ by maximizing the sum of the log-likelihoods across all graduating classes and student types:
\begin{equation}\label{eq:argmax1}
\hat{\theta}_1 = arg\max_{\theta_1} \sum_{s=1}^S\sum_{g=1}^G \sum_{t=1}^T \sum_{i=1}^{N_{tgs}} L_{itgs}\left(\theta_1\right).
\end{equation}
\subsection{Second step: Student's decision}
The log-likelihood of student $i$ in graduating class $g$ of high school $s$ taking the college-prep coursework is:
\begin{equation}\label{eq:studentlike}
L_{itgs}(\theta_2) = \mathbbm{1}\left\{a_{itgs}=1\right\}\log\left(\sigma_{itgs}\left(x_{gs};\hat{\theta}_1,\theta_2\right)\right) +  \left(1-\mathbbm{1}\left\{a_{itgs}=1\right\}\right)\log\left(1-\sigma_{itgs}\left(x_{gs};\hat{\theta}_1\theta_2\right)\right).
\end{equation}
In this equation, $a_{itgs}$ represents the (observed) coursework decision of student $i$ of type $t$ in graduating class $g$ in high school $s$, and $\sigma_{tgs}(\hat{\theta}_1,\theta_2)$ denotes the estimated conditional choice probability of a type-$t$ student in graduating class $g$ in high school $s$, based on the estimated parameter vector $\hat{\theta}_1$ and a candidate parameter vector $\theta_2$. For each candidate vector $\theta_2$, we solve for the vector of choice probabilities $\sigma_{itgs}\left(x_{gs};\hat{\theta}_1,\theta_2\right)$ that satisfy 

\begin{equation}\label{eq:studentccp}
    \sigma_{itgs}\left(x_{gs};\hat{\theta}_1,\theta_2\right) = \sigma_{itgs}\left(x_{gs},0; \theta_2\right) \left(1-\phi \left(x_{gs};\hat{\theta}_1\right)\right) + \sigma_{itgs}\left(x_{gs},1; \theta_2\right) \phi\left(x_{gs};\hat{\theta}_1\right) \text{ where }
\end{equation}
 $\sigma_{i t g s}\left(x_{gs} , b ;  \theta_2 \right) $ is the probability of student $i$ taking the college-prep coursework conditional on the teacher's decision $b$:

\begin{equation}\label{eq:br_type}
\sigma_{i t g s}\left(x_{gs} , b ;  \theta_2 \right) = \frac{\exp\left(x_{tgs}\beta + b\alpha + \kappa_g + \gamma_s +  \sum_{t^\prime } \lambda_{tt^\prime} w_{t t^\prime gs}\sigma_{t^\prime g s}\right)}{1+\exp\left(x_{tgs}\beta + b \alpha + \kappa_g + \gamma_s + \sum_{t^\prime} \lambda_{tt^\prime} w_{tt^\prime gs }\sigma_{t^\prime g s}\right)}.
\end{equation}
Therefore, this estimation step involves an iterative procedure that nests a fixed-point solution for the conditional choice probabilities inside maximum likelihood estimation \citet*{rust1987optimal}. We estimate the parameters $\theta_2$  by maximizing the sum of the log-likelihoods across all graduating classes and student types:
\begin{equation}\label{eq:argmax2}
\hat{\theta}_2 = arg\max_{\theta_2} \sum_{s=1}^S\sum_{g=1}^G \sum_{t=1}^T \sum_{i=1}^{N_{tgs}} L_{itg}(\theta_2).
\end{equation}

\section{Model Estimates}\label{results}
This section delves into the parameter estimates obtained from our analysis. Additionally, we present the marginal effects associated with the estimated parameters to facilitate a more intuitive interpretation of the results. To this end, we aggregate our sample of Senior cohorts to create one large, representative school. We assume an average value of the school-level unobserved heterogeneity. 

\subsection{Teacher's decision}

Table \ref{tab:mainresults}, Panel A, presents the estimated parameters regarding the teacher's decision to encourage students to apply to college. The results indicate that teachers are more likely to encourage female students, students with high grades in science/math courses, and students whose mothers attended college. However, they are less likely to encourage Black and Hispanic students than White students. This disparity decreases when Black or Hispanic students are in the majority in their class, especially for Hispanic students, where an increase in the Hispanic student population leads to a greater probability of Black and Hispanic students being encouraged to apply to college.\footnote{We normalize $\rho_{WW}=\rho_{BW}=\rho_{HW}=0$.} 

$$
\begin{pmatrix}
\rho_{WW} & \rho_{WB} & \rho_{WH} \\
\rho_{BW} & \rho_{BB} & \rho_{BH} \\
\rho_{HW} & \rho_{HB} & \rho_{HH} 
\end{pmatrix}
\equiv 
\begin{pmatrix}
0 & 0.81 & 0.36 \\
0 & 0.65 & 1.00^{**} \\
0 & 0.39 & 1.16^{***}
\end{pmatrix}
$$

\begin{table}[!t]\caption{Parameter Estimates}\centering\label{tab:mainresults}\fontsize{10}{10}\selectfont
\begin{threeparttable}
\begin{tabulary}{\linewidth}{lcc@{}}
\toprule
   & Estimate & s.e. \\ \midrule 
\addlinespace \multicolumn {3}{l}{\textbf{Panel A: Teacher's recommendation decision}} \\ \addlinespace
  Constant  &  -2.763\cellreff{$-2.763^{***}$}{tab1row19col1}   & 0.414 \\
  Black  &  -0.052\cellreff{$-0.052^{}$}{tab1row20col1}   & 0.104 \\
  Hispanic  &  -0.247\cellreff{$-0.247^{***}$}{tab1row21col1}   & 0.086 \\
  Female  &  0.168\cellreff{$0.168^{***}$}{tab1row22col1}   & 0.027 \\
  Course grade: A  &  0.415\cellreff{$0.415^{***}$}{tab1row23col1}   & 0.027 \\
  Mother attended college  &  0.551\cellreff{$0.551^{***}$}{tab1row24col1}   & 0.028 \\
  $\rho_{WB}$  &  0.810\cellreff{$0.810^{}$}{tab1row25col1}   & 0.680 \\
  $\rho_{BB}$  &  0.647\cellreff{$0.647^{}$}{tab1row26col1}   & 0.721 \\
  $\rho_{HB}$  &  0.386\cellreff{$0.386^{}$}{tab1row27col1}   & 0.705 \\
  $\rho_{WH}$  &  0.361\cellreff{$0.361^{}$}{tab1row28col1}   & 0.425 \\
  $\rho_{BH}$  &  1.008\cellreff{$1.008^{**}$}{tab1row29col1}   & 0.480 \\
  $\rho_{HH}$  &  1.158\cellreff{$1.158^{***}$}{tab1row30col1}   & 0.417 \\
\addlinespace \multicolumn {3}{l}{\textbf{Panel B : Student's graduation plan decision}} \\ \addlinespace
  Constant  &  -0.842\cellreff{$-0.842^{**}$}{tab1row1col1}   & 0.417 \\
  Black  &  1.237\cellreff{$1.237^{***}$}{tab1row2col1}   & 0.334 \\
  Hispanic  &  0.411\cellreff{$0.411^{**}$}{tab1row3col1}   & 0.193 \\
  Female  &  0.095\cellreff{$0.095^{***}$}{tab1row4col1}   & 0.027 \\
  Course grade: A  &  0.725\cellreff{$0.725^{***}$}{tab1row5col1}   & 0.028 \\
  Mother attended college  &  0.643\cellreff{$0.643^{***}$}{tab1row6col1}   & 0.030 \\
  Teacher encouraged for college  &  0.535\cellreff{$0.535^{***}$}{tab1row7col1}   & 0.030 \\
  $\lambda_{WW}$  &  3.561\cellreff{$3.561^{***}$}{tab1row8col1}   & 0.335 \\
  $\lambda_{BW}$  &  1.479\cellreff{$1.479^{***}$}{tab1row9col1}   & 0.498 \\
  $\lambda_{HW}$  &  2.367\cellreff{$2.367^{***}$}{tab1row10col1}   & 0.352 \\
  $\lambda_{WB}$  &  2.770\cellreff{$2.770^{***}$}{tab1row11col1}   & 0.692 \\
  $\lambda_{BB}$  &  0.755\cellreff{$0.755^{}$}{tab1row12col1}   & 0.828 \\
  $\lambda_{HB}$  &  1.804\cellreff{$1.804^{**}$}{tab1row13col1}   & 0.805 \\
  $\lambda_{WH}$  &  -1.123\cellreff{$-1.123^{}$}{tab1row14col1}   & 1.048 \\
  $\lambda_{BH}$  &  -2.858\cellreff{$-2.858^{***}$}{tab1row15col1}   & 1.063 \\
  $\lambda_{HH}$  &  -1.002\cellreff{$-1.002^{}$}{tab1row16col1}   & 0.974 \\
  \midrule Number of students  &   & 28,579  \\  

\\
\bottomrule
\end{tabulary}
\begin{tablenotes}[flushleft, online, normal, para]\scriptsize
\emph{Data source}: Texas Higher Education Opportunity Project (THEOP), Wave I, 2002
 \end{tablenotes}
\end{threeparttable}
\end{table}

The marginal effects of characteristics $x_{igs}$ on the probability of being encouraged to apply to college, $\Pr\left(b_{igs} =1 \mid x_{igs} = 1 \right) - \Pr\left(b_{igs} =1 \mid x_{igs} = 0 \right)$ can be summarized as follows: Black students are 3.04 percentage points less likely to be encouraged to apply to college than White students, Hispanic students are 3.54 percentage points less likely, and female students are 4.25 percentage points more likely. Students with an A grade in science/math coursework are 10.06 percentage points more likely to be encouraged, and students whose mothers attended college are 13.26 percentage points more likely.

The marginal effects of the racial composition on the teacher's decision, $\phi_{igs}(1-\phi_{igs})\rho_{ij}$, can be summarized as follows: An increase in the share of Black students from zero to one\footnote{An increase in the share of Black students from zero to one can be interpreted as the student being moved from a White-only school to a Black-only school.} leads to a 16.91 percentage point increase in college encouragement for White students, a 14.20 percentage point increase for Black students, and a 7.58 percentage point increase for Hispanic students. When the share of Hispanic students increases from zero to one, White students experience a 7.54 percentage point increase in college encouragement, Black students experience a 22.13 percentage point increase, and Hispanic students experience a 22.71 percentage point increase.

\subsection{Student's decision}

Panel B of Table \ref{tab:mainresults} presents the estimated parameters related to the student's decision to take the college-prep coursework. The results indicate that female students, those with higher course grades, and students whose mothers attended college are likelier to choose college-prep coursework. Additionally, students who receive encouragement from teachers to apply for college are more inclined to select college-prep coursework. Moreover, Black and Hispanic students are more likely to take college-prep coursework.

Three distinct patterns of heterogeneity in social incentives emerge from the analysis. Firstly, White students demonstrate the strongest social incentives for selecting the college-prep coursework, with a preference for their same-race and Black classmates choosing it. Secondly, Black and Hispanic students are more likely to opt for college-prep coursework when they anticipate their White classmates selecting it. The social incentives derived from their same-race classmates' decisions are less pronounced. Thirdly, Hispanic students are more likely to choose the college-prep coursework when they expect their Black classmates to choose it. Conversely, Black students are less inclined to select the college-prep coursework when they anticipate their Hispanic classmates opting for it.\footnote{To ensure the validity of the estimation results, we verify the assumption stated in Assumption (\ref{as:2}). This assumption establishes that Equation (\ref{eq:br_type}) is a contraction mapping, thereby guaranteeing the existence of a unique equilibrium. In order to test this assumption, we evaluate the condition $\frac{1}{N_{gs}-1}\sum_{j \in \mathcal{N}{gs} \setminus \{i\}} \mid \lambda{ij} \mid \leq 4 \forall g \in \{1,\ldots,G\}, s \in \{1,\ldots, S\}$ using the estimated parameters obtained from the model, and verify that it holds.}
$$
\begin{pmatrix}
\lambda_{WW} & \lambda_{WB} & \lambda_{WH} \\
\lambda_{BW} & \lambda_{BB} & \lambda_{BH} \\
\lambda_{HW} & \lambda_{HB} & \lambda_{HH} 
\end{pmatrix}
\equiv 
\begin{pmatrix}
3.56^{***} & 2.77^{***} & -1.12 \\
1.48^{***} & 0.75 & -2.86^{***} \\
2.37^{***} & 1.80^{**} & -1.00
\end{pmatrix}
$$

In the representative school, Black students are 2.08 percentage points less likely to take college-prep coursework than White students. Hispanic students are 6.76 percentage points less likely than White students and 4.68 percentage points less likely than Black students. Conversely, female students are 2.45 percentage points more likely to take college-prep coursework. Students who excelled in science/math courses have a 17.35 percentage point higher probability, while students with college-educated mothers have a 17.13 percentage point higher probability. Additionally, students who were encouraged to apply to college have a 15.18 percentage point higher probability of taking the college-prep coursework.

White students are 70.77 percentage points more likely to take the college-prep coursework when  their belief about their White classmates' decision to take it increases from zero to one.\footnote{An increase in the probability of a White student taking the college-prep coursework from zero to one can be interpreted as the student being moved from a school where the probability White students taking the college-prep coursework is zero to a class where this probability is one.} Black students exhibit a 15.70 percentage point increase in probability when they expect their Black classmates to select the college-prep coursework. Conversely, Hispanic students are 59.45 percentage points less likely to choose college-prep coursework when they perceive their Hispanic classmates' inclination toward it.

Furthermore, Black students are 30.78 percentage points more likely to take the college-prep coursework when they anticipate their White classmates' decision. In contrast, they are 59.45 percentage points less likely to take the college-prep coursework when the probability of their Hispanic classmates taking this coursework increases from zero to one. Hispanic students, on the other hand, are 70.77 percentage points more likely to choose the college-prep coursework when they anticipate their White classmates selecting it and 55.05 percentage points more likely when they expect their Black classmates to select it.

\section{Implications of desegregation}\label{counterfactual}

Integration policies implemented by school districts in the United States aim to address disparities in student outcomes and promote diversity \citep{TCFIntegrationReport}. These policies consider factors like socioeconomic status and race to determine student assignments, often involving changes in attendance boundaries or lottery-based enrollment \citep{Tractenberg2016, LearnedMiller2016}. Given the significant impact of these resource-intensive policies on individuals' lives, it is important to evaluate their potential implications.

In this section, we conduct a policy simulation exercise that offers valuable insights for policymakers aiming to improve student outcomes through school integration. We emphasize three key aspects that highlight the significance of this exercise in examining the implications of integration policies. First, we utilize the entropy index proposed by \citet{theil1972statistical} as a measure of segregation, enabling us to quantify the concentration or dispersion of student enrollment patterns across schools. Second, we address the complexity of student reassignment by treating it as a non-linear optimization problem, strategically redistributing students to generate alternative entropy index values. Finally, with the new distribution of students and estimated model parameters, we determine the probabilities of teachers' recommendations and students' coursework decisions under varying levels of segregation, revealing the impact of segregation on teachers' and students' behaviors.

\subsection{Measuring segregation: Entropy Index}

The \citet{theil1972statistical} index is particularly suitable for capturing how the diversity across individual schools diverges from the overall diversity at the state level, making it a valuable choice for our study. Furthermore, it can handle more than two groups, which aligns with the focus of our investigation. Additionally, the index adheres to the principle of transfers, accurately reflecting the effects of reassigning individuals and providing realistic insights into the impact of changes in school composition.

The \citet{theil1972statistical} index quantifies the diversity of a state's racial composition through an entropy score. The score is computed based on the sahres of students from each racial group in the state:
\begin{equation}\label{eq:entropyscorestate}
\bar{H} = -\sum_{r=1}^R  \ln \left(p_r^{p_r}\right),
\end{equation}
where $p_r$ represents the share of students belonging to racial group $r$ in the state. The maximum value of $\bar{H}$ for three racial groups is 1.099, indicating equal representation. 

The school-specific entropy score captures the diversity within each school. Our analysis treats each graduating class in the Senior cohort of the THEOP sample as an individual school. The school-specific entropy score is computed by summing the logarithm of the shares of students from each racial group, $p_{rg}$, within the school:
\begin{equation}\label{eq:entropyscoreschool}
H_g = -\sum_{r=1}^R \ln  \left(p_{rg}^{p_{rg}}\right).
\end{equation}

The \citet{theil1972statistical} index compares the racial composition of each school relative to the state-wide composition. It is calculated as the weighted average deviation of the school-level entropy scores from the state-wide entropy score:
\begin{equation}\label{eq:entropyindex}
H = -\sum_g p_g\frac{H_g - \bar{H}}{\bar{H}},
\end{equation}
where $p_g$ represents the share of students in school $g$ relative to the state-wide student population. This index provides a quantitative measure of segregation, considering both the diversity within individual schools and their representation in the overall student population. The entropy index ranges from zero to one, with zero indicating minimal segregation (when all schools have the same racial composition as the state) and one indicating maximal segregation (when schools consist exclusively of a single racial group).

\begin{figure}[!htbp]\caption{Entropy and Within-school Racial Composition}\label{fig:entropy}\centering
\begin{minipage}{0.45\linewidth}
\subcaption{Entropy = 0}
\includegraphics[width=\textwidth]{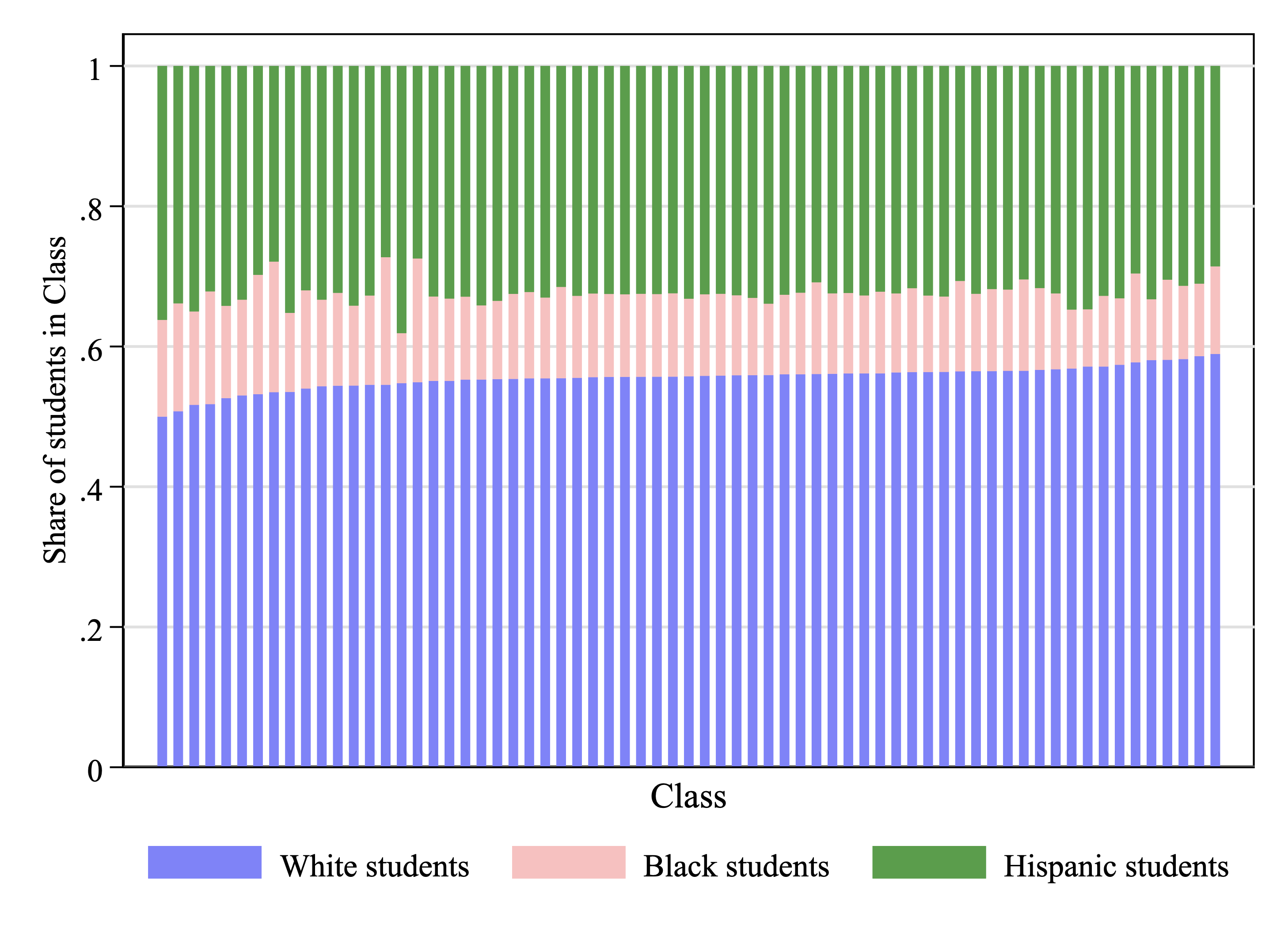}
\end{minipage}
\begin{minipage}{0.45\linewidth}
\subcaption{Entropy = 0.33}
\includegraphics[width=\textwidth]{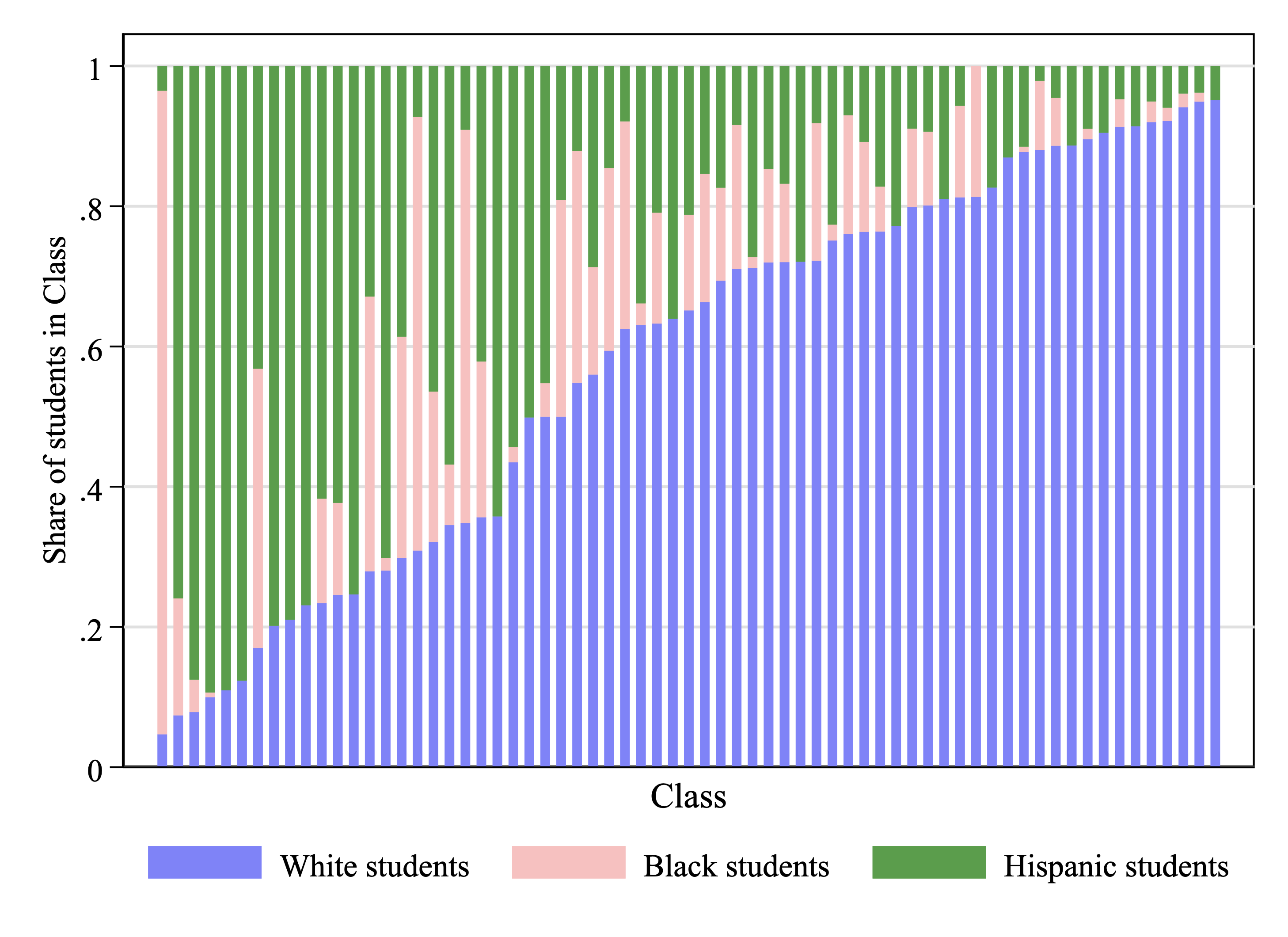}
\end{minipage}
\begin{minipage}{0.45\linewidth}
\subcaption{Entropy = 0.66}
\includegraphics[width=\textwidth]{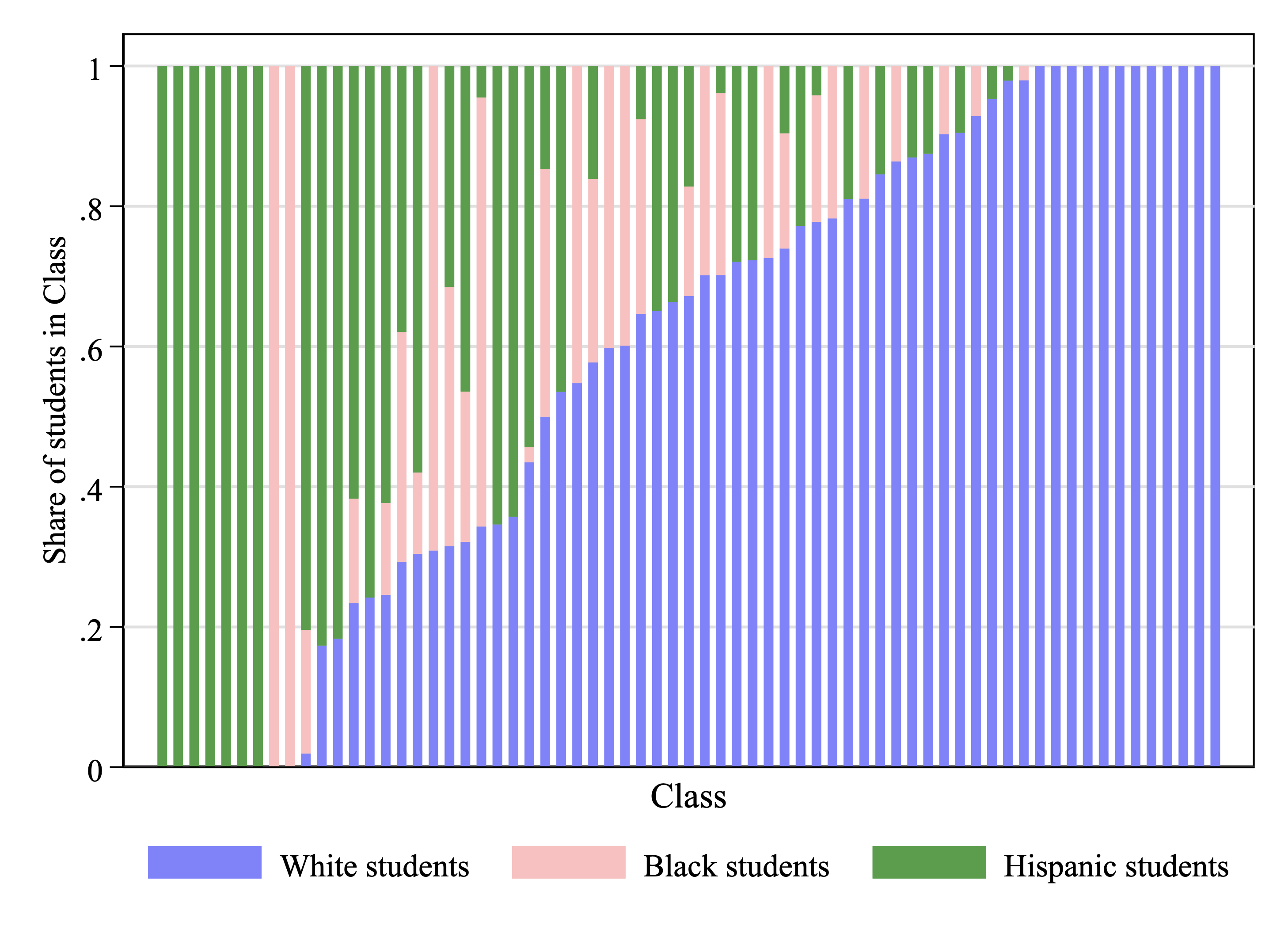}
\end{minipage}
\begin{minipage}{0.45\linewidth}
\subcaption{Entropy = 0.99}
\includegraphics[width=\textwidth]{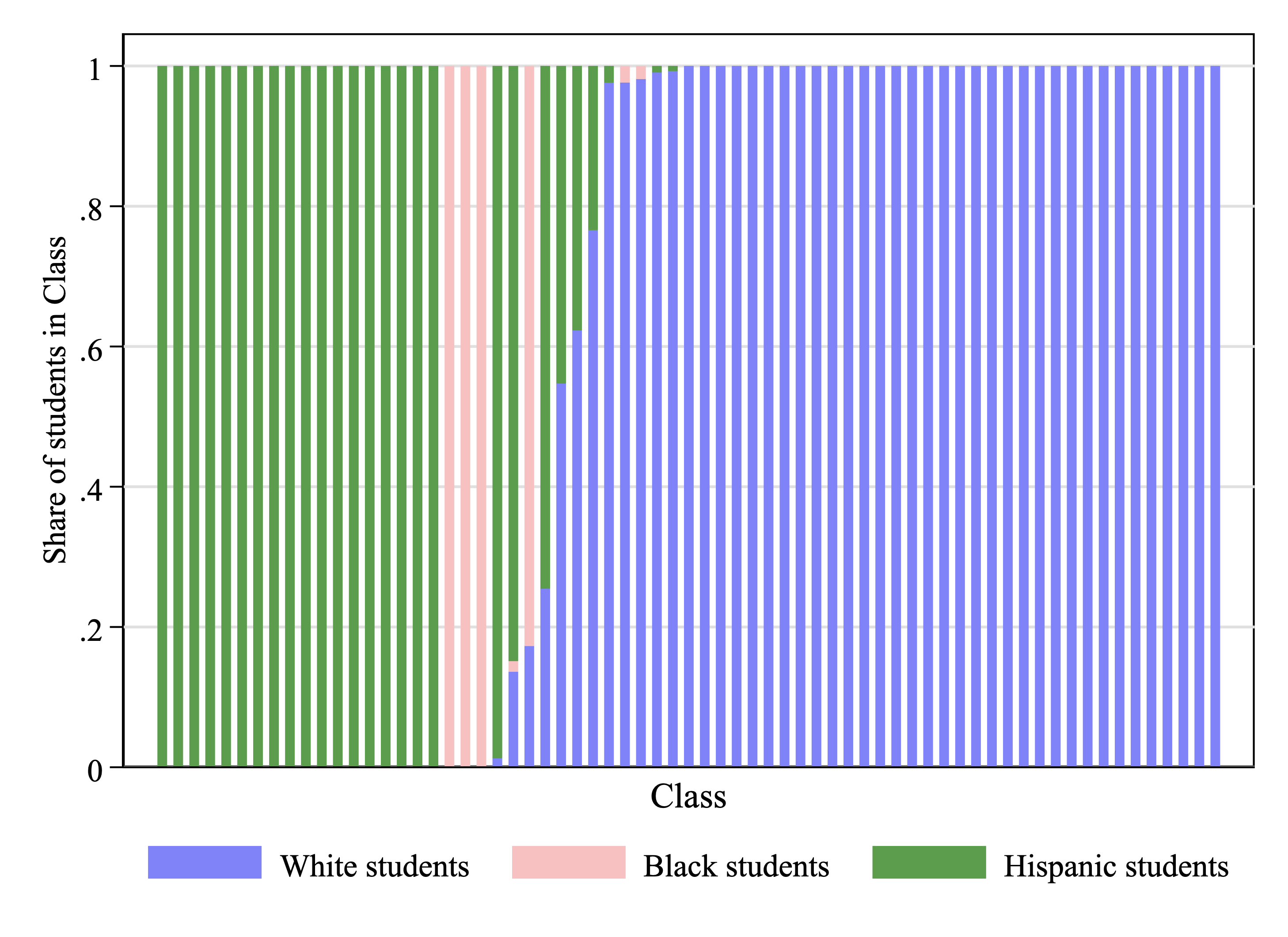}
\end{minipage}
\begin{minipage}{\linewidth}\scriptsize
\emph{Note:} All figures illustrate the distribution of the proportion of White, Black and Hispanic students across all schools simulated at different values of the entropy index.
 \end{minipage}
\end{figure}

\subsection{Student Reassignment}
To explore alternative segregation scenarios and achieve specific values of the entropy index, we formulate an optimization problem that involves redistributing students across schools. Each student is characterized by a set of observed characteristics, denoted by their type $t \in \{1,\ldots,T\}$. The objective is to find a school-specific student composition,  $\{p^{*}_{tg}\}_{t \in \{1,\ldots,T\},g\in\{1,\ldots,G}\}$, that closely resembles the observed student composition $\{\hat{p}_{tg}\}_\{t \in {1,\ldots,T\},g\in\{1,\ldots,G\}}$, in order to achieve a desired value of the entropy index, denoted as $H^{*}$. To accomplish this, we formulate an optimization problem:
\begin{equation}\label{eq:opt}
\begin{aligned}
&\min_{\{p_{tg}\}_{t\in\{1,\ldots,T\},g\in\{1,\ldots,G\}}} \frac{1}{G}\sum_{g=1}^G \sum_{t=1}^T \left(p_{tg}-\hat{p}_{tg}\right)^{2} \text{ subject to} \\
&\sum_{t\in\{1,\ldots,T\}}p_{tg}  = 1 \; \forall \; g\in\{1,\ldots,G\}\\
&\sum_{g=1}^G p_{tg}n_g = \hat{p}_tn \; \forall \; t\in\{1,\ldots,T\}\\
&\sum_g p_g\frac{H_g - \hat{H}}{\hat{H}} = H^{*} \text{ where } H_g = -\sum_{r=\{W,B,H\}} \ln  \left(\sum_{t:r_t = r}p_{tg}^{\sum_{t:r_t = r}p_{tg}}\right).
\end{aligned}
\end{equation}

The optimization problem aims to minimize differences between the constructed and observed school-specific student compositions while taking into account multiple constraints. Firstly, the shares of students of each observed type within each school must add up to one, maintaining the overall capacity of each school. Secondly, the population of each observed type across all schools should match the state-wide population. Finally, the constructed entropy index, which measures the level of segregation, should align with the target value.  To achieve these objectives, students are randomly assigned to new schools. Solving the optimization problem yields new school-level student compositions that closely resemble the observed compositions while achieving the desired level of segregation. Figure \ref{fig:entropy} illustrates within-school racial compositions ar different values of the entropy index, $H\in \{0,0.33,0.66,0.99\}$.

\subsection{Teachers' and Students' Decisions}
To evaluate the impact of student reassignment on teachers' recommendation and students' coursework decisions', we consider the alternative student composition, $\{p^{*}_{tg}\}_{t\in \{1,\ldots,T\},g\in \{1,\ldots,G\}}$, resulting from the student reassignment process. We determine students' and teachers' decisions in each school using the estimated utility parameters and the new student assignments. To isolate the effects of changes in student compositions across schools from the inherent differences across graduating classes and high schools, we assume constant average school-specific unobserved heterogeneity. This assumption allows us to focus specifically on the implications of student reassignment for teachers and students.

\begin{figure}[!htbp]\caption{Desegregation and Teacher's Recommendation and Student's Decisions}\label{fig:counterfactual}\centering
\begin{minipage}{0.45\linewidth}
\subcaption{Within-School Racial Composition and Teacher's Decision}
\includegraphics[width=\textwidth]{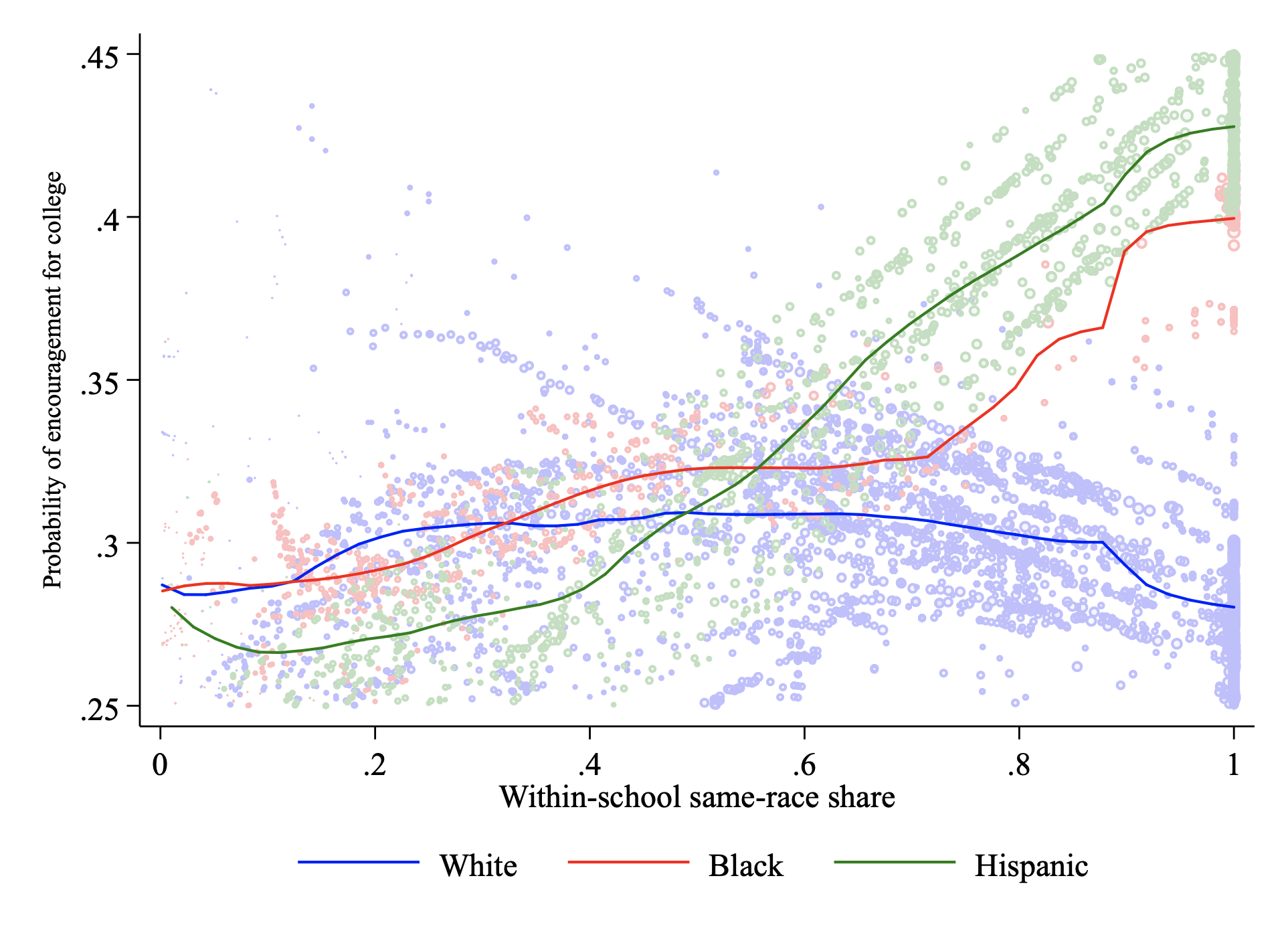}
\end{minipage}
\begin{minipage}{0.45\linewidth}
\subcaption{Between-School Desegregation and Teacher's Decision}
\includegraphics[width=\textwidth]{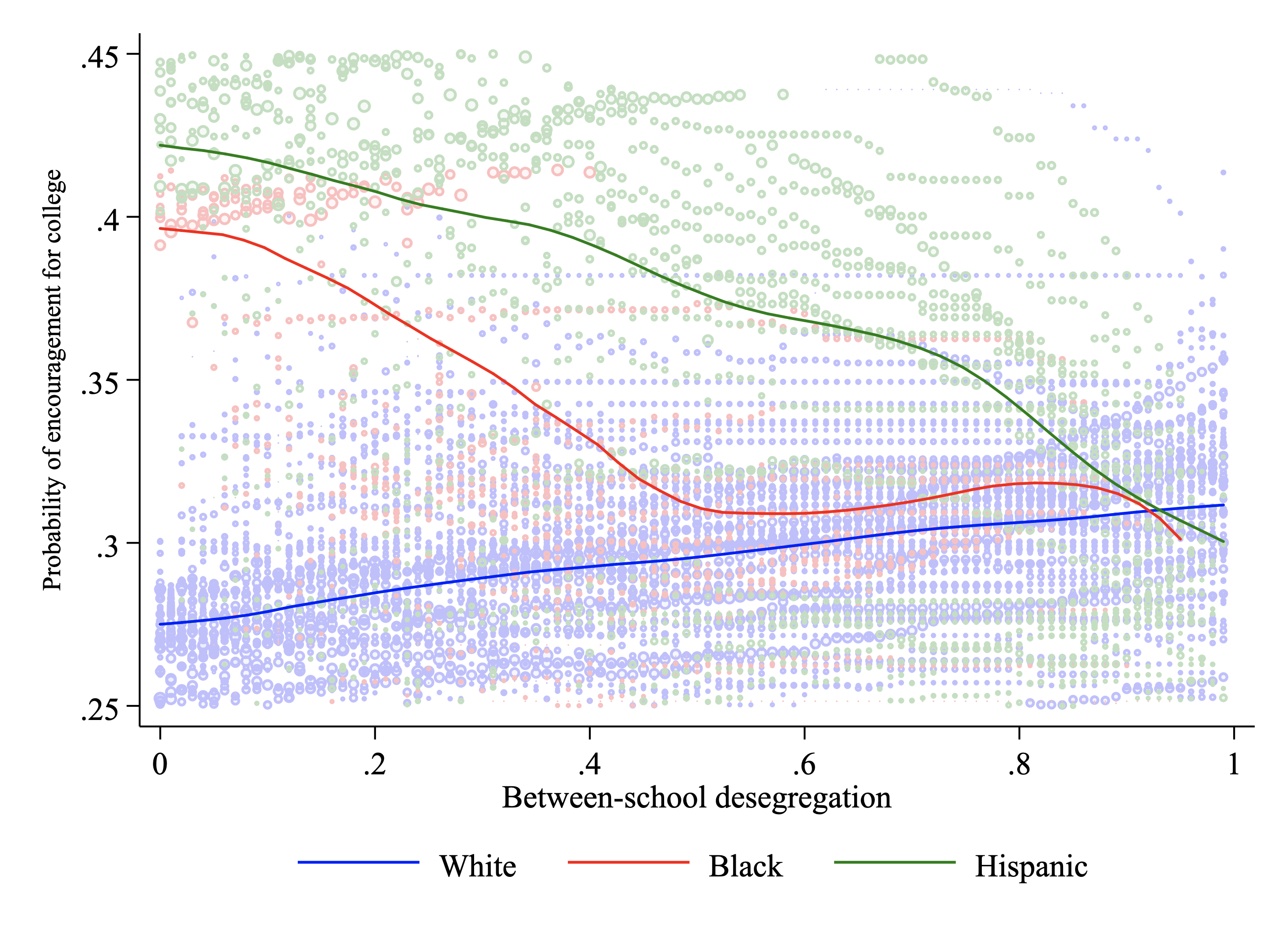}
\end{minipage}
\begin{minipage}{0.45\linewidth}
\subcaption{Within-School Racial Composition and Student's Decision}
\includegraphics[width=\textwidth]{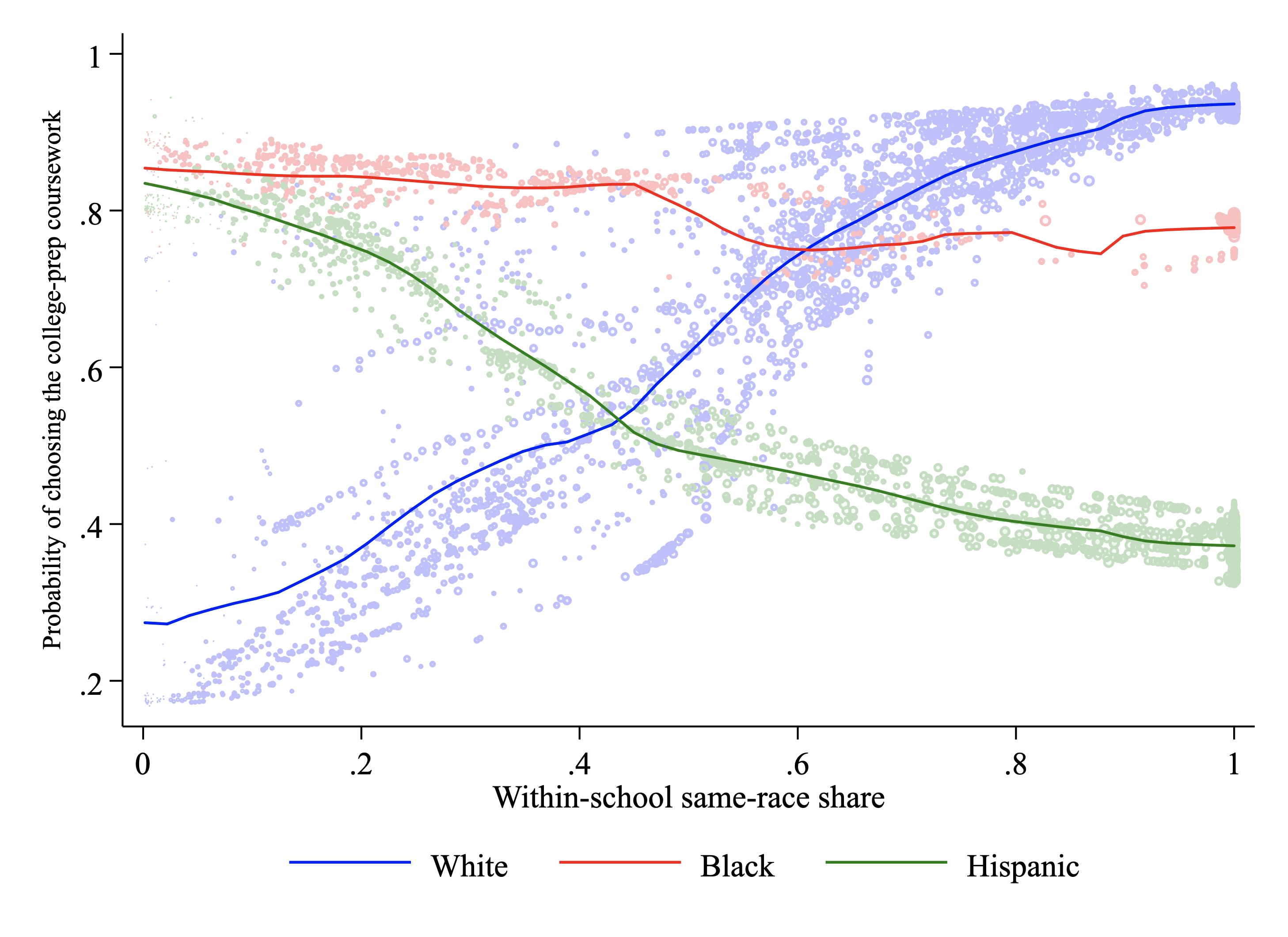}
\end{minipage}
\begin{minipage}{0.45\linewidth}
\subcaption{Between-School Desegregation and Student's Decision}
\includegraphics[width=\textwidth]{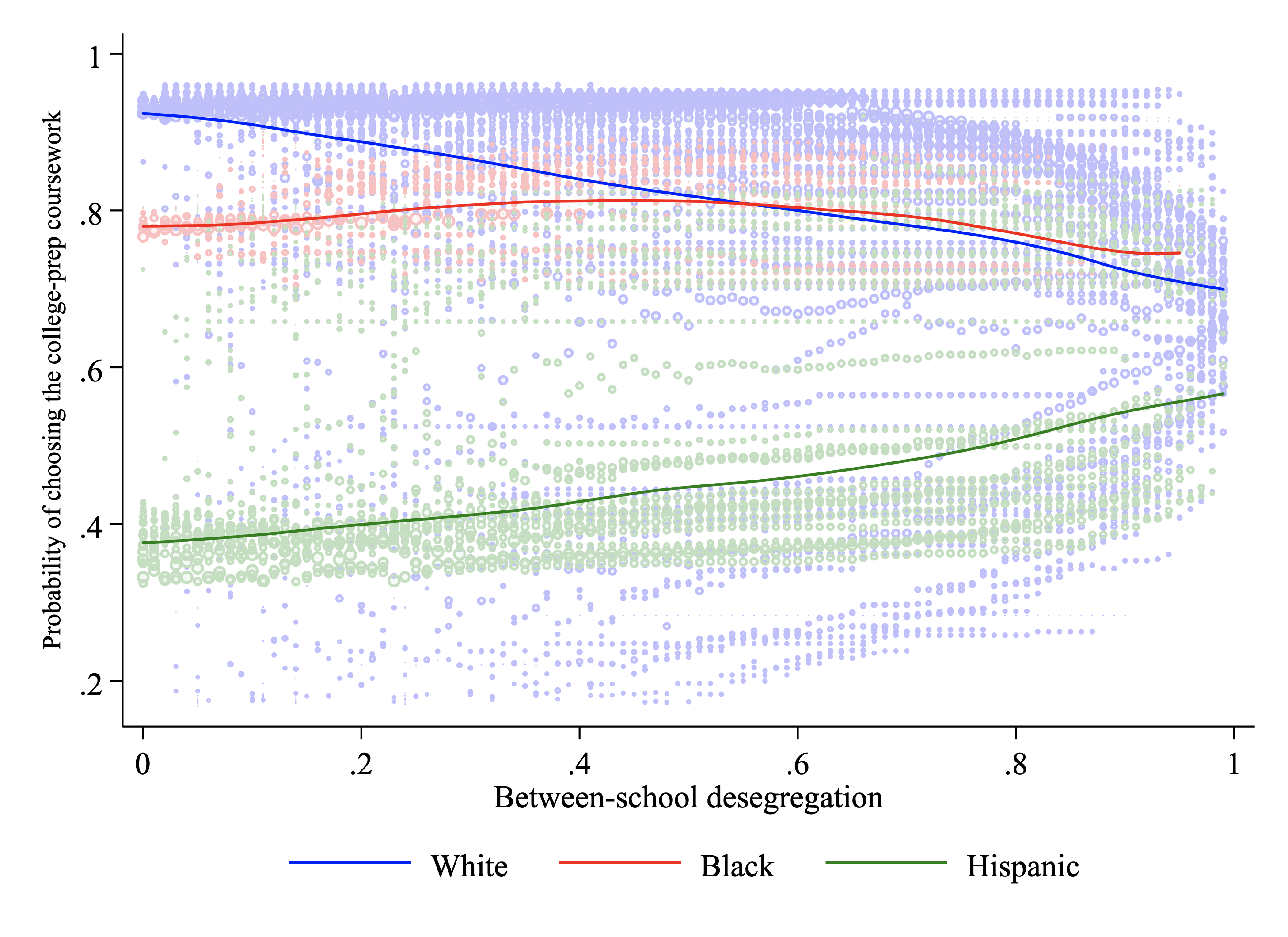}
\end{minipage}
\begin{minipage}{\linewidth}\scriptsize
\emph{Note:} Subfigure (a) displays the relationship between the probability of teachers encouraging students to apply to college and the share of same-race students within the school. Subfigure (b) shows the relationship between the probability of teachers encouraging students to apply to college and the level of desegregation between schools. Subfigure (c) illustrates the relationship between the probability of students choosing the college-prep coursework and the share of same-race students within the school. Subfigure (d) presents the relationship between the probability of students choosing the college-prep coursework and the level of desegregation between schools. All figures employ a non-parametric approach using an Epanechnikov kernel with a degree of 0 and a bandwidth of 0.05. Each marker in the figures represents a racial group from a specific school, with the size of the marker indicating the number of students within that racial group in that school.
 \end{minipage}
\end{figure}

Figure \ref{fig:counterfactual} presents the probabilities of teachers encouraging students to apply to college and students choosing the college-prep coursework at different levels of within-school share of same-race students and between-school desegregation. The analysis focuses on schools with predominantly White or Black students (with less than 10\% Hispanic students) when examining Black students' decisions and teachers' decisions regarding encouraging Black students. Similarly, for Hispanic students, the analysis is limited to schools with a predominant racial composition of White or Hispanic (with less than 10\% Black students). This restriction allows us to examine the impact of changes in between-school segregation and the share of same-race classmates relative to White students on teachers' and students' behaviors.

Subfigures (a) and (c) illustrate the relationship between the share of same-race students in the class and the probabilities of students being encouraged to apply to college and choosing the college-prep coursework. As the share of same-race students increases, both Black and Hispanic students are more likely to be encouraged to apply to college, but they are less likely to choose the college-prep coursework, with this pattern being more pronounced among Hispanic students.

Subfigures (b) and (d) depict how the probabilities of students being encouraged to apply to college and choosing the college-prep coursework vary with between-school desegregation, represented by $(1-H)$ where $H \in [0,1]$ is the entropy measure of between-school segregation. As between-school segregation decreases and schools more closely reflect the state-level racial composition, White students are more likely to be encouraged to apply to college but less likely to choose the college-prep coursework. Conversely, as between-school segregation decreases, Black and Hispanic students are less likely to be encouraged to apply to college but more likely to choose the college-prep coursework.

\section{Conclusion}\label{conclusion}
This research provides valuable insights into the intricate relationship between desegregation initiatives, institutional bias, and individual behavior in education. By employing a simultaneous-move game model and utilizing entropy to measure segregation, we provide a more comprehensive understanding of how between-school desegregation can influence disparities in school coursework by influencing both teachers' perceptions of student abilities and students' perceptions of each others' behaviors.

Through our analysis of high school student's decisions to take college-prep coursework and teachers' decisions regarding encouraging students to apply to college, we uncover notable heterogeneity in how teachers steer students towards college and how students sort into college-prep coursework. Our findings highlight that Black and Hispanic students are likelier to take college-prep coursework as between-school segregation decreases. However, we also find that a decrease in between-school segregation is associated with teachers being less likely to encourage Black and Hispanic students to apply to college. 

To address disparities in student outcomes, many school districts have implemented integration policies, including measures such as redrawing student attendance zones or lottery-based student reassignment. In order to support policymakers in making informed decisions regarding these policies, we propose a data-informed framework that involves constructing alternative levels of segregation by solving an optimization problem. This framework seeks to find alternative distributions of student composition across schools that closely resemble the observed distribution while considering the constraints imposed by available school resources and the demographic composition of the student population.

By offering an extensive analysis of the complexities surrounding educational equity and the impact of desegregation efforts on institutional bias and individual behavior, our research contributes to a deeper understanding of the consequences of desegregation initiatives. Given the substantial scale, resource requirements, and life-altering implications of school integration policies, our proposed methodology provides valuable insights for assessing the impact of these policies and striving for a more equitable education system.

\pagebreak
\bibliographystyle{chicago}

\end{document}